\documentclass[runningheads,a4paper]{llncs}

\usepackage{amssymb}
\usepackage{graphicx}

\usepackage{url}
\urldef{\mailsa}\path|amironov66@gmail.com|
\urldef{\mailsb}\path|anna.kramer, leonie.kunz, christine.reiss, nicole.sator,|
\urldef{\mailsc}\path|erika.siebert-cole, peter.strasser, lncs}@springer.com|    
\newcommand{\keywords}[1]{\par\addvspace\baselineskip
\noindent\keywordname\enspace\ignorespaces#1}

\begin{document}

\def\dey#1#2{#1 (#2)}
\def\deyc#1#2{#1 \cdot  #2}

\def\ral#1{\;\mathop{\longrightarrow}\limits^{#1}\;}
\def\bc{\begin{center}\begin{tabular}{l}}
\def\ec{\end{tabular}\end{center}}
\def\modn#1{\mathop{=}\limits_{#1}}
\def\modnop#1{\mathop{#1}\limits_{n}}
\def\rar{\mathop{\in}\limits_{r}}
\def\und#1{\mathop{=}\limits_{#1}}
\def\skobq#1{\langle\!| #1 |\!\rangle}
\def\redeq{\;\mathop{\approx}\limits^{r}\;}
\def\reduc{\;\mathop{\mapsto}\limits^{r}\;}
\def\pt{\;\mathop{+}\limits_{\tau}\;}
\def\sost{\begin{picture}(0,0)\put(0,3){\circle*{4}}
\end{picture}}
\def\sosto{\begin{picture}(0,0)\put(0,0){\circle*{4}}
\end{picture}}
\def\bi{\begin{itemize}}
\def\pa{\,|\,}
\def\oc{\;\mathop{\approx}\limits^{+}\;}
\def\p#1#2{(\;#1\;,\;#2\;)}
\def\mor#1#2#3{\by #1&\pright{#2}&#3\ey}
\def\ei{\end{itemize}}
\def\bn{\begin{enumerate}}
\def\en{\end{enumerate}}
\def\i{\item}
\def\a{\forall\;}

\def\l#1{\langle #1 \rangle}

\def\bigset#1#2{\left\{\by #1 \left| \by #2 \ey\right\}\ey\right.}
\def\p{\leftarrow}

\def\plongright#1{
  \begin{picture}(40,8)
  \put (-5,3){\vector(1,0){50}}
  \put (20,8){\makebox(1,1){$\scriptstyle #1$}}
  \end{picture} }

\def\plongleft#1{
  \begin{picture}(40,8)
  \put (45,3){\vector(-1,0){50}}
  \put (20,8){\makebox(1,1){$\scriptstyle #1$}}
  \end{picture} }

\def\pse#1#2{
  \begin{picture}(40,8)
  \put (-5,-5){\vector(2,-1){50}}
  \put (45,-5){\vector(-2,-1){50}}
  \put (-5,-12){\makebox(1,1)[r]{$\scriptstyle #1$}}
  \put (45,-12){\makebox(1,1)[l]{$\scriptstyle #2$}}
  \end{picture} }

\def\und#1{\mathop{=}\limits_{#1}}
\def\redeq{\;\mathop{\approx}\limits^{r}\;}
\def\reduc{\;\mathop{\mapsto}\limits^{r}\;}
\def\oc{\mathop{\approx}\limits^{+}}
\def\sost{\begin{picture}(0,0)\put(0,0){\circle*{4}}
\end{picture}}
\def\bi{\begin{itemize}}
\def\pa{\,|\,}
\def\oo{\;\mathop{\approx}\limits^{c}\;}
\def\p#1#2{(\;#1\;,\;#2\;)}
\def\mor#1#2#3{\by #1&\pright{#2}&#3\ey}
\def\ei{\end{itemize}}
\def\bn{\begin{enumerate}}
\def\en{\end{enumerate}}
\def\i{\item}
\def\bigset#1#2{\left\{\by #1 \left| \by #2 \ey\right\}\ey\right.}
\def\p{\leftarrow}
\def\buffer{{\it Buffer}}
\def\eam{\mathbin{{\mathop{=}\limits^{\mbox{\scriptsize def}}}}}
\def\be#1{\begin{equation}\label{#1}}
\def\ee{\end{equation}}
\def\re#1{(\ref{#1})}

\def\bn{\begin{enumerate}}
\def\en{\end{enumerate}}
\def\bi{\begin{itemize}}
\def\ei{\end{itemize}}
\def\i{\item}
\def\c#1{\left\{\begin{array}{lllll}#1\end{array}\right\}}
\def\d#1{\left[\begin{array}{lllll}#1\end{array}\right]}
\def\b#1{\left(\begin{array}{lllll}#1\end{array}\right)}
\def\ra#1{\;\mathop{\to}\limits^{#1}\;}
\def\leqd{\;\mathop{<}\limits_{2}\;}
\def\diagrw#1{{
  \def\normalbaselines{\baselineskip20pt \lineskip3pt \lineskiplimit3pt }
  \matrix{#1}}}

\def\blackbox{\vrule height 7pt width 7pt depth 0pt}
\def\pu#1#2{
\mbox{$\!\!\begin{picture}(0,0)
\put (-#1,-#2){\line(1,0){#1}}
\put (-#1,-#2){\line(0,1){#2}}
\put (#1,#2){\line(-1,0){#1}}
\put (#1,#2){\line(0,-1){#2}}
\put (-#1,#2){\line(1,0){#1}}
\put (-#1,#2){\line(0,-1){#2}}
\put (#1,-#2){\line(-1,0){#1}}
\put (#1,-#2){\line(0,1){#2}}
\end{picture}$}
}

\def\pright#1{
  \begin{picture}(20,8)
  \put (-5,3){\vector(1,0){30}}
  \put (9,8){\makebox(1,1){$\scriptstyle #1$}}
  \end{picture} }

\def\by{\begin{array}{llllllllllllll}}
\def\ey{\end{array}}

\mainmatter  

\title{A graph model of message passing processes}



\author{Andrew M. Mironov%
%
}

\institute{Moscow State University\\
\mailsa\\
}

\maketitle

\begin{abstract}
In the paper we consider a graph model of 
message passing
processes and 
present 
a method verification 
of message passing
processes. 
The method is 
illustrated by an example of a verification of
sliding window protocol.
\end{abstract}


\section{Introduction}

The problem of formal representation and verification of discrete processes is
one of the most important problems in computer science. There are several
approaches to this problem, the main of them are: CCS and 
$\pi$-calculus \cite{1}, \cite{2},
CSP and its generalizations \cite{3}, temporal logic and model checking \cite{4}, Petri nets
\cite{5}, process algebras 
\cite{6}, communicating finite-state machines \cite{7}.

In the present paper we introduce a new model of discrete processes, which
is a synthesis of Milner's model of processes \cite{1} and the model of communicating
finite-state machines \cite{7}. Discrete processes are represented in our model as graphs, edges of which are labelled by  operators. These operators consist of internal actions and communication actions. Proofs
of correctness of processes are represented by sets of formulas, associated with
pairs of states of analyzed processes. This method of verification of processes
is a synthesis of Milner's approach related on the concept of an observational
equivalence \cite{1} and Floyd's inductive assertion method \cite{8}. 
For a simplification of an analysis of processes
we introduce a simplification operation on processes. With use this operation it is possible
to reduce a complexity of verification of processes. 
We illustrate an advantage of
the proposed model and the verification method on the example of verification
of a two-way sliding window protocol.

\section{Motivation, advantages of the proposed approach and
its comparison with other works}

\subsection{Motivation of the proposed approach}

The main disadvantage of modern methods
of verification  of discrete processe
is their large complexity.
More precisely,
\bi
\i the main disadvantage of verification methods based on model checking approach is a high computational complexity related to the state explosion problem,
and
\i disadvantages of methods based on theorem proving approach are related
with a high complexity of
construction of corresponging theorems and their proofs, and also with an 
understanding of these proofs.
\ei
For example, in recent paper \cite{9} 
a complete presentation of proofs of theorems
related to verification of two-way sliding window protocol takes a few dozen
pages of a complex mathematical text.

The main motivation for the proposed approach to modeling and verification
of discrete systems by checking of observational 
equivalence of corresponded message passing
processes 
is to simplify and make more obvious the following aspects of modeling and analysis of discrete systems:
 representation of mathematical models of analyzed systems,
 construction of proofs of correctness of the systems, and
 understanding of these proofs by any who is not a strong expert in the
mathematical theory of verification of discrete systems.

\subsection{Advantages of the proposed approach}

The proposed mathematical model of message passing 
processes with allows to
construct such mathematical models of analysed systems that are very similar
to an original description of these systems on 
any  imperative programming 
language. In section \ref{sec8} we give an example of such model that corresponds to a
C-program describing a sliding window protocol using go back $n$ (the program
was taken from book \cite{12}, section 3.4.2).

The main advantage of the proposed approach
 is a possibility to use a simplification operation of models
of analyzed systems,
that  allows essentially simplify the problem of verification of these models.
In section \ref{sec8} we present a result of such simplification for the above model of a
sliding window protocol: this model can be simplified to a model with only one
state. 
It should be noted also 
that the simplified models allow more clearly understand main features of analyzed systems, and facilitate a construction
of correctness proofs for analyzed systems.

If an analyzed property of a system
has the form of a behavior which is described by some process, for example, in the case when
\bi
\i  an analyzed system is a network protocol, and
\i a property of this system is a description of an external behavior of this protocol
(related to its interaction with a
higher-level protocol)
\ei
then a proof of a correctness of such system in this model is a set of formulas
associated with pairs of states, the first of which is 
a state of the analyzed system, and
the second is a state of a 
a process which describes a property 
of the analyzed system.

In section \ref{sec8} we give an example of such proof, which is a small set of simple
formulas. These formulas can be naturally derived from a simplified model of an analyzed protocol.

Another advantage of the proposed approach is a possibility to verify systems
with unbounded sets of states. One of examples of such systems is the above
sliding window protocol using go back $n$.

\subsection{Comparison with other works}

In this section we present an overview of papers related to verification of message
passing systems, which are most relevant to the present paper.

The paper \cite{9} deals with modeling and manual verification in the process
algebraic language $\mu$CRL. Authors use the theorem prover PVS to formalize
and to mechanically prove the correctness of a protocol using selective repeat (a
C-program describing this protocol is presented in section 3.4.3 of the book \cite{12}).
The main disadvantage of this work is a large complexity of proofs of theorems
related to verification of this protocol. This protocol can be verified more simply
with use of the approach proposed in the present paper.

There are a lot of works related to verification of systems with message
passing based on temporal logic and model checking approach. Most relevant
ones to the present paper are 
\cite{13}, \cite{14}, \cite{15}, 
\cite{16}, \cite{17}, \cite{18}, 
\cite{19}. The most
deficiency of all of them is restricted abilities: these methods allow verify only
finite state systems.

Among other approaches it should be noted  approaches with use of first
order logic and assertional verification: 
\cite{20}, \cite{21}, and approaches with use of
process algebra: 
\cite{22}. 
The most deficiency of these approaches is
a high complexity of construction of proofs of correctness of analyzed systems.

\section{Auxiliary concepts}

\subsection{Terms}

We assume that there are given
a set ${\cal X}$ of {\bf variables}, 
a set ${\cal D}$ of {\bf values}, 
a set ${\cal C}$ of {\bf constants},  and 
a set ${\cal F}$ of {\bf function symbols}.
Any constant from  ${\cal C}$ is interpreted by
a value from  ${\cal D}$, and 
any function symbol
 from ${\cal F}$ is interpreted by
an operation on ${\cal D}$.

We assume that ${\cal C}$
 contains constants 0 and 1, 
and ${\cal F}$
contains boolean function symbols $\wedge,\vee,\to$,
which correspond to 
standard boolean operations
on $\{0,1\}$.

The set ${\cal E}$ of {\bf terms}
is defined in  the standard way. 
Variables and constants are terms. 
Other terms have the form
$f(e_1,\ldots, e_n)$, where
$f\in {\cal F}$, and
$e_1,\ldots, e_n$ are terms.
For each $e\in {\cal E}$ 
a set of all variables occurring in $e$
is denoted by $X_e$.

If $X\subseteq {\cal X}$, 
then a {\bf valuation} of variables of $X$ 
is a correspondence $\xi$, 
that associates each variable $x\in X$ 
with a value $x^\xi\in {\cal D}$.
We denote by the record $X^\bullet$
the set of all valuations 
of variables from $X$. 
For each $e \in {\cal E}$, 
each $X \supseteq X_e$ 
and each $\xi \in X^\bullet$ 
the record $e^\xi$ 
denotes an object called 
a {\bf value} of $e$ on $\xi$ 
and defined in the standard way. 
We assume that terms $e_1$ and $e_2$ 
are equal iff 
$\forall\,\xi\in (X_{e_1}\cup X_{e_2})^\bullet\;\;
e_1^\xi=e_2^\xi$.

A term $e$ is a {\bf formula} if 
$\forall\,\xi\in  X_e^\bullet$
the value $e^\xi $ is 0 or 1. 
The set of all formulas
is denoted by ${\cal B}$. 
The symbols $\top$ and $\bot$ 
denote
true and false formula respectively. 
We shall write 
formulas of the form $\wedge(b_1,b_2)$,
$\vee(b_1,b_2)$,
etc. in a more familiar form 
$b_1\wedge b_2$, $b_1\vee b_2$, etc. 

\subsection{Atomic operators}
\label{oe}

We assume that there is given 
a set ${\cal N}$, whose elements
are considered as names of objects 
that can be sent or received by processes.

An {\bf atomic operator (AO)} is an
object $o$ of one of three
forms presented below. 
Each pair ($o,\xi$), where
$o$ is an AO, and $\xi$ is a valuation 
of variables occurred in $o$,
corresponds to an action
$o^{\xi}$, informally defined below.
\bn
\i
An {\bf input} is an  AO of the form 
$\alpha?x$, where $\alpha\in  {\cal N}$
and $x \in {\cal X}$.
An action
$(\alpha?x)^{\xi}$
is a receiving from another process
an object named $\alpha$, with
a message attached to this object,
this message is assigned to the variable $x$.

\i An {\bf output } is an AO of the form
$\alpha ! e$, where
$\alpha\in  {\cal N}$ and $e \in {\cal E}$. 
An action $(\alpha ! e)^{\xi}$ 
is a sending to another process
an object named
$\alpha$, 
to which a message $e^\xi$ is attached.

\i An {\bf assignment} is an AO
of the form 
$x: = e$, where $x \in {\cal X}$,
$e \in {\cal E}$.
An action
$(x: = e)^{\xi}$ 
is an assigning
the variable $x$
with the value $e^\xi$.
\en

Below we use the following notations.
\bi
\i For each AO $o$
the record $X_{o}$ 
denotes the set of
all variables occurred in $o$.
\i If  $e\in {\cal E}$, and
$o$ is an assignment, then
the record $\dey{o}{e}$ 
denotes a term defined as follows: let
$o$ has the form  $(x:=e')$, then
$\dey{o}{e}$ is obtained from $e$
by a  replacement of all occurrences 
of the variable $x$ by the term $e'$.
\i If $o$ is an assignment, and
 $\xi\in X^\bullet$,
where $X_o\subseteq X\subseteq {\cal X}$, 
then the
record $\deyc{\xi}{o}$
denotes a valuation from $X^\bullet$,
defined as follows:
let $o=(x:=e)$, then
$x^{\deyc{\xi}{o}}= e^\xi$
and $\forall\,y\in X\setminus \{x\}\;\;
y^{\deyc{\xi}{o}}= y^\xi$.
\ei

It is easy to prove that if
$o$ is an assignment and $e\in {\cal E}$,
then for each $\xi\in X^\bullet$,
where $X_o\cup X_e\subseteq X\subseteq {\cal X}$,
the equality
$\dey{o}{e}^\xi=e^{\deyc{\xi}{o}}$ holds.
This equality is  proved by
an induction on the structure of the term $e$.

\subsection{Operators}
\label{concatt}

An {\bf  operator} is a record $O$ of the form
$b\,[o_1, \ldots, o_n]$,
where $b$ is a formula called
a {\bf  precondition} of $O$
(this formula
will be denoted as $\l{O}$),
and
$o_1, \ldots, o_n$ is a sequence
of AOs  (this sequence
will be denoted as $[O]$),
among which there is at most 
one input or output.
The sequence $[O]$ may be empty  ($[\,]$).

If $[O]$ contains an input (or an output)
then $O$ is called an  {\bf input  operator}
(or an  {\bf output  operator}), 
and in this case the  record
$N_O$ denotes a name  occurred in $O$.
If $[O]$ does not contain 
inputs and outputs,
then we call $O$ an {\bf  internal operator}.

If $\l{O}=\top$, then such precondition
can be omitted in a notation of $O$.

Below we use the following notations.
\bn
\i For each operator $O$
a  set of
all variables occurred in $O$
is denoted by $X_{O}$.

\i If $O$ is an operator, and
$b\in {\cal B}$, 
then the record  $\deyc{O}{b}$
denotes an object,
which either is  a formula or
is not defined.
This object
is defined recursively as follows.
If $[O]$ empty, then
$\deyc{O}{b}\eam \l{O}\wedge b$.
If
$[O]=o_1,\ldots, o_n$, where  $n\geq 1$, then
we shall denote by the record
$O\setminus o_n$ an operator obtained 
from $O$ by a removing of its last AO,  and
\bi
\i if $o_n=\alpha?x$, then
$\deyc{O}{b}\eam \deyc{(O\setminus o_n)}{b}$,
if  $x\not\in X_b$, and 
  is undefined
otherwise
\i if $o_n=\alpha!e$,
then $\deyc{O}{b} \eam
   \deyc{(O\setminus o_n)}{b}$
\i if $o_n = (x:=e)$, then $\deyc{O}{b} \eam
   \deyc{(O\setminus o_n)}{\dey{o_n}{b}}$.
     \ei

\i If $O$ is an internal operator, and
 $\xi\in X^\bullet$,
where $X_O\subseteq X\subseteq {\cal X}$, 
then the record  $\deyc{\xi}{O}$
denotes a valuation from $X^\bullet$,
defined as follows:
if $[O]$ is empty, then
$\deyc{\xi}{O}\eam \xi$,
and if
$[O]=o_1,\ldots, o_n$, where $n \geq 1$, 
then
$\deyc{\xi}{O}\eam \deyc{(\deyc{\xi}{(O\setminus o_n)})}{o_n}$.
\en

It is easy to prove that if
$O$ is internal and $b \in {\cal B}$,
then for each  $\xi\in X^\bullet$,
where $X_O\cup X_b\subseteq X\subseteq {\cal X}$,
such that  $\l{O}^\xi=1$,
the equality
$(\deyc{O}{b})^\xi=b^{\deyc{\xi}{O}}$ holds.
This equality is
proved by an 
induction on a lenght of $[O]$.

\subsection{Concatenation of operators}

Let $O_1$ and $O_2$ be operators, and 
at least one of them is internal.

A {\bf concatenation} of $O_1$ and $O_2$
is an object denoted by the record
$O_1 \cdot O_2$, that either is 
operator or is undefined.
This object is defined iff
$\deyc{O_1}{\l{O_2}}$ is defined, and 
in this case
$O_1\cdot O_2\eam
(\deyc{O_1}{\l{O_2}})[[O_1], [O_2]]$.
It is easy to prove that 
\bi
\i if operators $O_1, O_2$ and formula $b$ 
are such that objects in both
sides of the equality 
$\deyc{(O_1\cdot O_2)}{b} = 
\deyc{O_1}{(\deyc{O_2}{b})}$
are defined, then this equality holds, and
\i if operators $O_1, O_2, O_3$ are 
such that
all objects in both sides of the equality
$(O_1\cdot O_2)\cdot O_3 = O_1\cdot (O_2\cdot O_3)$
are defined, then this equality holds.
\ei

\section{Message passing
processes}

\subsection{A  concept of a message passing process}

A {\bf message passing
process}
(also called more briefly 
a {\bf process}) is a graph
$P$ of the form 
\be{process}P=(S_P, s^0_P,  T_P, I_P)\ee
components of which have the following meanings.
\bi
\i $S_P$ is a  set of nodes of  $P$, 
which are called
{\bf states} of the process $P$.
\i  $s^0_P \in S_P$ is an {\bf initial state} of the 
process $P$.
\i $T_P$ is a  set of edges of the graph $P$,
which are called
{\bf transitions}, each transition from $T_P$
  has the form
$s_1\ra{O}s_2$, 
where $s_1, s_2 \in S_P$, and $O$ 
is an 
operator, 
which is a label of this edge.
\i $I_P\in {\cal B}\setminus \{\bot\}$ is a 
{\bf  precondition} of the process $P$.
\ei

A transition $s_1\ra{O}s_2$ is called an
{\bf input}, an {\bf output}, 
or an {\bf internal} transition, if $O$ is
an input operator, an  output operator, 
or an  internal operator, respectively.

For each process $P$
\bi
\i the record $X_P$ denotes the set
consisting of \bi\i all variables occurred in 
any of the transitions from $T_P$, or
in $I_P$, and 
\i a variable $at_P$, which is  
not occurred in $I_P$,
and in  transitions from $T_P$,
the set of values of $at_P$ is $S_P$ 
\ei
\i the record
$\l{P}$ denotes the 
formula  $(at_P=s^0_P) \wedge I_P$.
\ei

For each transition $t\in T_P$
the records $O_t$, $\l{t}$, $start(t)$ and $end(t)$
denote an operator, a formula and states 
defined as follows:
if $t$ has the from 
$s_1\ra{O}s_2$,
then  $$O_t\eam O,\;\;
\l{t}\eam (at_P=s_1)\wedge \l{O},\;\;
start(t)\eam s_1,\;\; end(t)\eam s_2.$$
If $t$ is an input or an output, then the record
$N_t$ denotes  the name $N_{O_t}$.

A set $X^s_P$ of {\bf essential variables} of 
$P$ is a smallest (w.r.t. inclusion)
set satisfying the following conditions.
\bi
\i $X^s_P$ contains all variables
contained in preconditions
and outputs in operators $O_t$,
where $t\in T_P$.
\i If $P$ contains an AO 
$x: = e$ and $x\in X^s_P$,
then $X^s_P$ contains all variables
occurred in $e$.
\ei

\subsection{Actions of processes} 

An {\bf action of a process }  (or, briefly, an {\bf action})
is a record of one of the following three forms.
\bi
\i  $\alpha?d$, where $\alpha\in {\cal N}$ and 
$d\in {\cal D}$.
An action of this form is called a {\bf receiving}
of an object named $\alpha$
with the attached message $d$.

\i  $\alpha!d$, where 
$\alpha\in {\cal N}$ and $d\in {\cal D}$.
An action of this form is called a {\bf sending}
of an object named $\alpha$
with the attached message $d$.

\i $\tau$.
An action of this form is called a {\bf silent action}.
\ei

A set of all actions is  denoted by ${\cal A}$.

\subsection{An execution of a process}

An {\bf  execution} of a process 
\re{process} is a walk on the graph $P$
starting from $s^0_P$, with an execution of 
AOs occurred in labels of traversed edges.
At each step $i\geq 0$ of this walk
there is defined a current state $s_i\in S_P$
and a current valuation $\xi_i \in X_P^\bullet$. 
We assume that
$s_0= s^0_P$,
$\l{P}^{\xi_0} = 1$, 
and for each step $i$ of this walk 
$at_P^{\xi_i} = s_i$.

An execution of $P$ on step $i$
is described informally as follows.
If there is no transitions in $T_P$ starting at $s_i$, 
then $P$ terminates, otherwise
\bi
\i $P$ selects a transition $t\in T_P$, 
such that
$\l{t}^{\xi_i}=1$,  and
if $t$ is an input or an output, then 
at the current moment 
$P$ can receive or send respectively
an object named $N_t$
(i.e. at the same moment there is
another process that can send to $P$
or receive from $P$ respectively
an object named $N_t$).
If there is no such transition,
then $P$ suspends until
at least one such transition will appear, 
and after resumption its execution
$P$ selects one of such transitions,
\i after a sequential execution of all AOs
occurred in the operator $O_t$ of the 
selected transition $t$, $P$
moves to the state $end(t)$. 
\ei

An execution of each AO $o$ occurred in $[O_t]$
consists of a performing of an action $a\in {\cal A}$
and a replacement the current valuation 
$\xi$ on a valuation $\xi'$, 
which is considered as a current
valuation after an execution of the AO $o$.
An execution of an AO $o$ is as follows:
\bi
\i if $o = \alpha ? x$, then $P$ 
performs an action of the form $\alpha?d$,
and $x^{\xi'} \eam d$,
$\forall\,y \in X_P \setminus \{x\}\quad
y^{\xi'} \eam y^{\xi}$
\i if $o = \alpha ! e$, then 
$P$  performs the action 
$\alpha!(e^{\xi})$, and
$\xi'\eam \xi$
\i if $o =  (x := e)$, then 
$P$  performs $\tau$, 
and $x^{\xi'} \eam e^{\xi}$,
$\forall\,y \in X_P \setminus \{x\}\quad
y^{\xi'} \eam y^{\xi}$.
\ei

\section{Operations on processes} 
\label{erqwr2343242}

In this section we define some operations on processes which can be used 
for a construction of 
complex processes from simpler ones. 
These operations are generalizations of
corresponded operations on processes
defined in Milners's Calculus 
of Communicating Systems \cite{1}.

\subsection{Parallel composition}

The operation of parallel composition is used for 
building processes, 
composed of several communicating subprocesses. 

Let $P_i=(S_i, s^0_i,  T_i, I_i)\;\;(i=1,2)$ 
be processes, such that
$S_1\cap S_2=\emptyset$ and
$X_{P_1}\cap X_{P_2}=\emptyset$.
A {\bf parallel composition} 
of $P_1$ and $P_2$
is a process
$P=(S_, s^0,  T, I)$, 
where
$S \eam S_1\times S_2$,
$s^0\eam (s^0_1, s^0_2)$,
$I\eam I_1\wedge I_2$,
and $T$ consists of the following transitions:
\bi
\i for
each transition
$s_1\ra{O}s'_1$
of the process $P_1$, and
each state $s$ of $P_2$
the process $P$ 
has the transition
$(s_1,s)\ra{O}(s'_1,s)$

\i for
each transition
$s_2\ra{O}s'_2$
of the process $P_2$, and
each state $s$ of the process $P_1$
the process $P$ 
has the transition
$(s,s_2)\ra{O}(s,s'_2)$

\i for each pair of transition of the form 
$\left\{
\by
s_1\ra{O_1}s'_1\;\in T_{P_1}\\
s_2\ra{O_2}s'_2\;\in T_{P_2}\ey
\right.$
where one of the operators
$O_1$, $O_2$ 
has the form
$(O'_1\cdot[\alpha?x])\cdot
O''_1$,
and another operator has the form
$(O'_2\cdot [\alpha!e])\cdot O''_2$,
the process  $P$ 
has the transition
$(s_1,s_2)\ra{O}(s'_1,s'_2)$, where
$\l{O} = \l{O_1}\wedge\l{O_2}$ and
$[O]=
\Big((O'_1\cdot
O'_2)\cdot
[x:=e]\Big)\cdot
(O''_1\cdot
O''_2)
$.
\ei

A parallel composition of $P_1$ and $P_2$
is denoted by the record $P_1\pa P_2$.

If $S_1\cap S_2\neq\emptyset$ or
$X_{P_1}\cap X_{P_2} \neq\emptyset$,
then before a construction of the process
$P_1\pa P_2$ it is necessary to replace
states and variables occuring in both processes
on new states or variables respectively.

 For any tuple $P_1,P_2,\ldots, P_n$ 
 of processes their 
  parallel  composition $P_1\pa \ldots\pa P_n$
 is defined as the process 
 $((P_1\pa P_2)\pa \ldots)\pa P_n$.

\subsection{Restriction}

Let $P=(S, s^0,  T, I)$ be a process,
and $L$ be a subset of the set ${\bf N}$.

A {\bf restriction} of $P$ with respect to $L$
is the process
$P\setminus L = (S, s^0, T', I)$
which is obtained from $P$ by removing of those
transitions that have labels with the names from $L$, i.e.
$T' \eam \bigset{(\diagrw{s&\pright{O}&s'})
\in R}{[O]=[\;],\;\;\mbox{or}  \; N_O\not\in L}$.

\subsection{Renaming}

The last operation is 
called a {\bf renaming}:
for any mapping 
$f: {\cal N}\to {\cal N}$
and any process $P$ the record 
$P[f]$ denotes a process 
which is called a renaming of $P$ and 
is obtained from  $P$ by
changing of names occurred in $P$:
any name $\alpha$ occurred in $P$
is changed on $f(\alpha)$.

If the mapping $f$ acts 
non-identically only on the names
$\alpha_1,\ldots, \alpha_n$, 
and maps them to the names
$\beta_1,\ldots, \beta_n$
respectively, then the process $P[f]$ 
can be denoted also as
$P[\beta_1/\alpha_1,\ldots, \beta_n/\alpha_n]$.

\section{Realizations of processes}

\subsection{Realizations of AOs and 
sequences of AOs}
\label{xio}

A {\bf realization of an AO $o$} is a  triple 
$(\xi, a, \xi')$, such that
\bi
\i $\xi,\xi' \in X^\bullet$,
where $X_o\subseteq  X\subseteq {\cal X}$,
and $a\in {\cal A}$
\i if $o=\alpha?x$, then 
$a=\alpha?(x^{\xi'})$ and
$\forall\,y\in X\setminus \{x\}\quad
y^{\xi'}=y^\xi$
\i if $o=\alpha!e$, then $a=\alpha!(e^\xi)$
and $\xi'=\xi$
\i if $o=(x:=e)$, then $a=\tau$ and
$\xi'=\deyc{\xi}{o}$.
\ei

Let $o_1,\ldots, o_n$ be a sequence 
of AOs which contains at most one input or output.
A {\bf realization of 
$o_1,\ldots, o_n$}
is a triple  $(\xi, a, \xi')$, such that
\bi
\i $\xi,\xi' \in X^\bullet$, where $X\subseteq {\cal X}$ 
and $a\in {\cal A}$
\i if $n=0$, then $\xi'=\xi$ and $a=\tau$, otherwise
there exists a sequence
\be{sdfsadfsadfsa}
(\xi_0, a_1, \xi_1),\;
(\xi_1, a_2, \xi_2),\;\ldots,
(\xi_{n-1}, a_n, \xi_n)\ee where
$\xi_0=\xi$, $\xi_n=\xi'$, $\forall\,i=1,\ldots, n\;\;
(\xi_{i-1}, a_i, \xi_i)$ is a 
realization of  $o_i$,  and $a=\tau$,
if each $a_i$ in \re{sdfsadfsadfsa} is equal to $\tau$, 
otherwise $a$ coincides with that  $a_i$, 
which is different from $\tau$. \ei

\subsection{Realization of transitions}
\label{realper}

Let $P$ be a process  of the form
\re{process}, and $t\in T_P$.

A {\bf realization of $t$}
is a triple
 $(\xi_1,a,\xi_2)$, where
$\xi_1, \xi_2\in X_P^\bullet$ and
$a\in {\cal A}$, such that 
$\l{t}^{\xi_1}=1$ and
$(\xi_1\cdot(at_P:=end(t)),a,\xi_2)$ is a 
realization of $[O_t]$.

The following properties hold.
\bi
\i If a transition $t$ is internal or is an output, 
then for each $\xi\in X_P^\bullet$, such that $\l{t}^\xi=1$,
there exist a unique $\xi'\in X_P^\bullet$ and
a unique  $a\in {\cal A}$, such that
 $(\xi, a, \xi')$ is a realization of $t$. 
We shall denote such $\xi'$ by $\deyc{\xi}{t}$.
\i If a transition  $t$ is an input, 
then for each  $\xi\in X_P^\bullet$,
such that  $\l{t}^\xi = 1$, and each $d\in {\cal D}$
there exists a unique $\xi'\in X_P^\bullet$, such that
 $(\xi, N_t?d, \xi')$  is a realization of $t$. 
We shall denote such $\xi'$ by $\deyc{\xi}{t^d}$.
\ei

\subsection{Realizations of processes}
\label{defpr}

A {\bf realization of a process $P$}
is a graph $P^r$
having the following components.
\bi
\i The set $S_ {P}^r$ of vertices of $P^r$ 
is the disjoint union $X_P^\bullet \cup\{P^{0}\}$.
\i The set $T_ {P}^r$ of edges of $P^r$
   consists of the following edges:
\bi\i for each realization $(\xi_1, a, \xi_2)$ of any $t\in 
 T_P$ the graph $P^r$ 
has an edge from $\xi_1$ to $\xi_2$ with a label $a$,
and 
\i for each $\xi \in X_P^\bullet$,
such that $\l{P}^{\xi} = 1$, and
each edge of $P^r$
from $\xi$ to $\xi'$ with a label $a$
the graph $P^r$ has 
an edge from $P^{0}$ to $\xi'$ with a label $a$.
\ei\ei

We shall use the following notations:
for any pair $v,v'$ of vertices of $P^r$
\bi
\i the record $v_1\ra{a} v_2$ denotes an edge
from $v_1$ to $v_2$ with a label $a$
\i $v\ra{\tau^*} v'$ means that
either $v=v'$ or
$\exists\,v_0, v_1, \ldots, v_n:$
$\forall\,i=1,\ldots, n$ the graph
$P^r$ has an edge
$v_{i-1}\ra{\tau}v_i$, and
$v_0 = v$, $v_n = v'$.
\i  $v\ral{\tau^*a\tau^*} v'$ (where 
$a\in {\cal A}$)
means that
$\exists\,v_1, v_2:$  the graph $P^r$ 
 has an edge $v_1\ra{a}v_2$, 
and
$v\ral{\tau^*} v_1$, $v_2\ral{\tau^*} v'$.
\ei

\section{Observational equivalence of processes}

\subsection{A concept of observational equivalence of processes}\label{definition}

Processes $P_1$ and $P_2$ are said to be
{\bf observationally equivalent} 
if $P^r_1$ and $P^r_2$ 
are observationally equivalent in Milner's sense
 \cite{1}, i.e. there exists
$\mu\subseteq S_{P_1}^r \times S_{P_2}^r$,  
such that
\bn
\i $(P_1^0, P_2^0)\in \mu$
\i if $(v_1, v_2)\in \mu$ and $v_1\ra{\tau}v'_1$,
then
$\exists\,v'_2: v_2\ra{\tau^*}v'_2,\;
(v'_1, v'_2)\in \mu$, \\
if  $(v_1, v_2)\in \mu$ and $v_2\ra{\tau}v'_2$,
then
$\exists\,v'_1: v_1\ra{\tau^*}v'_1,\;
(v'_1, v'_2)\in \mu$
\i if $(v_1, v_2)\in \mu$ and $v_1\ra{a}v'_1$,
  $a\neq \tau$,
then 
$\exists\,v'_2: v_2\ral{\tau^*a\tau^*}v'_2,\;
(v'_1, v'_2)\in \mu$, \\
if $(v_1, v_2)\in \mu$ and $v_2\ra{a}v'_2$,
 $a\neq \tau$,
then 
$\exists\,v'_1: v_1\ral{\tau^*a\tau^*}v'_1,\;
(v'_1, v'_2)\in \mu$
\en

The record $P_1\approx P_2$ means that
$P_1$ and $P_2$ are 
observationally equivalent.

A lot of problems related 
to verification of discrete systems
can be reduced to the problem 
to prove that $P_1\approx P_2$,
where the process $P_1$
is a model of a system being analyzed,
and $P_2$ is a model of some property
of this system. 
In section \ref{sec8}
we consider an example of a proof
that $P_1\approx P_2$, 
where $P_1$ is a model of the 
sliding window protocol, 
and $P_2$ is a model of its external behavior.

\subsection{A method of a proof of observational 
equivalence of processes}
\label{theor}

In this section we present a method of 
a proof of observational equivalence of processes.
This method is based on theorem \ref{th1}. 
To formulate and prove this theorem, 
we introduce auxiliary concepts and notations.

\bn
\i Let $P$ be a process, and $s,s'\in S_P$. 
A {\bf composite transition (CT)} from $s$ to $s'$
is a sequence $T$ of transitions of $P$ of the form
\be{l7}
s=s_0\ra{O_1}s_1,\quad
s_1\ra{O_2}s_2,\quad\ldots\quad
s_{n-1}\ra{O_n}s_n=s'
\ee
such that there is at most one input or output
operator among $O_1,\ldots, O_n$, and 
there are defined all concatenations in the 
expression
\be{sdsdfsdfs444}(\ldots(O_1\cdot O_2)\cdot\ldots)\cdot O_n\ee

Sequence \re{l7} may be empty, in this case $s=s'$.
If CT $T$ is not empty and has the form \re{l7}, 
then the record $O_T$ denotes
a value of the expression \re{sdsdfsdfs444}.
If CT $T$ is empty, then $O_T \eam [\,] $.

We shall use for CTs the same concepts 
and notation as for ordinary transitions  
($start(T)$, $end(T)$, $N_T$ etc.). 
A CT $T$ is said to be an input, an output, 
or an internal iff $O_T$  is an input operator, 
an output operator, or an internal operator, respectively.

A concept of a realization of a CT is defined 
by analogy with the concept of a realization of a transition (see section \ref{realper}). 
This concept  has properties similar to 
properties of a realization of a transition,
in particular:
\bn
\i if  a CT $T$ is internal or is an output,
then for each $\xi\in X_P^\bullet$,
such that $\l{T}^\xi=1$,
there is a unique $\xi'\in X_P^\bullet$ and a unique
$a\in {\cal A}$, such that
 $(\xi, a, \xi')$ is a realization of $T$,
we shall denote such $\xi'$
 by the record $\deyc{\xi}{T}$
\i if  a CT  $T$ is an input, 
then for each $\xi\in X_P^\bullet$,
such that $\l{T}^\xi = 1$,
and each $d\in {\cal D}$
there is a unique $\xi'\in X_P^\bullet$, such that
 $(\xi, N_T?d, \xi')$ is a realization of 
$T$, we shall denote such $\xi'$
by the record $\deyc{\xi}{T^d}$.
\en

\i If $b$ and $b'$ are formulas, then 
the record $b\leq b'$ is a brief notation of the 
proposition that the formula $b\to b'$ is true.

\i If $O_1, O_2$  are operators, andè
$b\in {\cal B}$, then the record
$\deyc{(O_1,O_2)}{b}$
denotes a formula defined 
by a recursive definition presented below.
In this definition we use records of 
the form $O\setminus o$ and $o(b)$,
which denote an operator and a formula
respectively, defined in section  \ref{concatt}.

Let $[O_1]=o_1,\ldots, o_n$ and  
$[O_2]=o'_1,\ldots, o'_m$, then the formula
\be{fgvfdsgdsfgdsfgd}\deyc{(O_1,O_2)}{b}\ee 
is defined as follows:
\bn
\i $\l{O_1}\wedge \l{O_2}\wedge b$, if $n=m=0$
\i $\deyc{(O_1\setminus o_n, O_2)}{o_n(b)}$, 
if $o_n$ is an assignment
\i $\deyc{(O_1, O_2\setminus o'_m)}{o'_m(b)}$, if 
$o'_m$ is an assignment
\i \label{peremz}
$\deyc{((O_1\setminus o_n), (O_2\setminus o'_m))}
{b(z/x,z/y)}$, 
if $o_n=\alpha?x,\;o'_m=\alpha?y$,
and $b(z/x,z/y)$ is a formula obtained from $b$
replacing all occurrences of
$x$ and $y$ on a fresh variable
$z$ (i.e. $z$ is not occurred in $O_1$,
$O_2$ and $b$)
\i $\deyc{((O_1\setminus o_n), (O_2\setminus o'_m))} {((e_1=e_2)\wedge b)}$, 
if $o_n=\alpha!e_1$ and $o'_m=\alpha!e_2$
\i $\bot$, otherwise.\en\en

\refstepcounter{theorem}
{\bf Theorem \arabic{theorem}\label{th1}}

Let
$P_i=(S_{P_i}, s^0_{P_i}, T_{P_i}, \l{{P_i}})\;\;(i=1,2)$ 
be processes such that 
$S_{P_1} \cap S_{P_2} = \emptyset$
and $X_{P_1} \cap X_{P_2} = \emptyset$.
Then $P_1\approx P_2$, if 
there exist a set
$\{b_{s_1s_2}\mid
s_i\in S_{P_i}\;(i=1,2)\}$
of formulas with variables from 
$(X_{P_1}\cup X_{P_2})\setminus\{at_{P_1},
at_{P_2}\}$, such that
\bn
\i \label{sdfsa332}
$\l{P_1}\wedge \l{P_2}\leq 
b_{s^0_{P_1}s^0_{P_2}}$
\i \label{sdfsa33}
$\forall\,(s_1\ra{O}s'_1)\in T_{P_1},\;
\forall\,s_2\in S_{P_2}$
there exists a set $\{s_2\ra{T_i}s_2^i\mid i\in\Im\}$ 
of CTs of 
   $P_2$ 
such that 
$b_{s_1s_2} \wedge \; \l{O}
\leq
\bigvee\limits_{i\in\Im}
\deyc{(O, O_{T_i})}
{b_{s'_1s^{i}_2}}
$
\i $\forall\,(s_2\ra{O}s'_2)\in T_{P_2},\;
\forall\,s_1\in S_{P_1}$
there exists a set $\{s_1\ra{T_i}s_1^i\mid i\in\Im\}$   
of CTs of  $P_1$  such that 
$b_{s_1s_2} \wedge \; \l{O}
\leq
\bigvee\limits_{i\in\Im}
\deyc{(O_{T_i}, O)}
{b_{s^{i}_1s'_2}}
$
\en

\section{Simplification of processes}
\label{simp}

\subsection{A concept of a simplification of processes}

The concept of a simplification of processes
is intended to reduce the problem 
of verification of processes.

A {\bf simplification} of a process $P$
is a sequence of 
transformations of this process, 
each  of which is performed
according to one of  the rules set out below.
Each of these transformations 
(except the first) is performed on the result 
of previous transformation. 
A {\bf result} of a simplification
is a result of last of  these transformations. 

Simplification rules are defined as follows.
Let $P$ be a process.
\begin{description}
\i[Rule 1 (removing of states).]$\;$\\
If $s\in S_P\setminus \{s^0_P\}$, and 
\bi
\i $s_1\ra{O_1}s$, $\ldots$,
$s_n\ra{O_n}s$ are all transitions incoming to $s$ 
\i $s\ra{O'_1}s'_1$, $\ldots$,
$s\ra{O'_m}s'_m$
are all transitions outgoing from $s$,
and if all these transitions are internal, then
$\l{O'_i}\wedge \l{O'_j}=\bot$
if $i\neq j$
  \i $s\not\in\{s_1,\ldots,
     s_n,s'_1,\ldots,s'_m\}$
  \i  $\forall\,i=1,\ldots, n,\; 
     \forall\,j=1,\ldots, m\quad
\exists\,O_i \cdot O'_j$ 
\ei
then $s$ and all transitions related to $s$
are removed from $P$, and the transitions 
$s_i\ral{O_i\cdot O'_j}s'_j$
(where $i=1,\ldots, n,\;
		 j=1,\ldots, m$)
are added to $P$.
\i[Rule 2 (fusion).]$\;$\\
If $P$ has a pair of transitions of the form
$s_1\ra{O}s_2$,
  $s_1\ra{O'}s_2$,
and $[O]=[O']$,
then this pair is replaced by a  transition
$s_1\ral{b[O]}s_2$, where
$b=\l{O}\vee \l{O'}$.

\i[Rule 3 (elimination of unessential assignments).]$\;$\\
If $P$ has an AO  $(x:=e)$, where 
$x\not\in X^s_P$,
  then this AO is removed from $P$.
\end{description}

\refstepcounter{theorem}
{\bf Theorem \arabic{theorem}\label{th2}}.
If $P'$ is a result of simplification of $P$,
then $P'\approx P$.

\section{An example: verification of a
sliding window protocol}
\label{sec8}

In this section we present an 
example of use of theorem \ref{th1}
for a verification of a
sliding window protocol.
This protocol ensures 
a transmission of messages from one
agent to another through a medium, in which messages may get distorted or
lost. In this section we consider a two-way sliding window protocol, in which the
agents can both send and receive messages 
from each other. We do not present
here a detail explanation of this protocol, a reader can find it in section 3.4.2 of
the book \cite{12} (a protocol using go back $n$).

\subsection{A structure of the protocol}

The protocol is a system consisting of  interacting components,
including
\bi
\i  components that perform a formation, sending, receiving and processing of
messages (such components are called {\bf agents}, and messages sent from one
agent to another, are called {\bf frames}), and
\i a medium, through which frames are forwarded (such a medium is called a
{\bf channel}).
\ei
A detailed description of the components 
and relation between them is represented 
in the Appendix.

\subsection{Specification}

External actions of the above protocol (i.e. actions which are related to its communication with a network level) have the form $In_1 ? d$, $In_2 ? d$, $Out_1 ! d$ 
and
$Out_2 ! d$. Assume that we take into account only external actions $In_1 ? d$ and
$Out_2 ! d$, and ignore other its external actions (i.e. we consider a transmission 
only in one direction: from the left to the right). We
would like to prove that such behavior is equivalent to a behavior of a process
$B_{n-1}$, which is called ``a FIFO buffer which can hold at most $n-1$ frames'', and
is defined as follows:
\bi\i  
variables of $B_{n-1}$ are
\bi
\i an array $(x[0],\ldots, x[n-1])$, 
elements of which have the same type as a
type of frames in the above protocol, and
\i variables $r, s, u$, values of which belong to ${\bf Z}_n$, and have the following
meaning: at every moment
\bi
\i  a value of $u$ is equal to a number of frames in the buffer
\i  values $r$ and $s$ can be interpreted as lower and upper bounds of a
part of the array $x$, which stores the received frames, which has not
yet been issued from the buffer
\ei
\ei
\i $B_{n-1}$ has one state and 2 transitions with labels
   $$
   \by
   (u<n-1)\,[In?x[s],\;s:=s\modnop{+}1,\;u:=u+1]\\
   (u>0)\;[Out\,!\,x[r],\;r:=r\modnop{+}1,\;u:=u-1]
      \ey
   $$
   where $\forall\,i\in \{0,n-2\}\;\;i\modnop{+}1\eam
     i{+}1$ and $(n-1)\modnop{+}1 \eam 0$
\i  initial condition is $r = s = u = 0$.
\ei

A process that describes a behavior of the protocol with respect to the above
specific point of view (where we ignore actions of the form $In_2 ? d$ and $Out_1 ! d$)
is constructed as a parallel composition of the processes corresponded to components of this procotol, with elimination of atomic operators
 related to ignored communications. 

\subsection{Verification}

With use of the simplification operations from section 
\ref{simp}, we can transform
the process corresponded to the protocol
(with elimination of atomic operators
which are corresponded
to ignored actions) to a process   $P$
with only one state and with 
transitions labelled by the following operators:
\bi
\i $(w<n-1)\;[In?x[s], \; M_1:=M_1\cdot \varphi(x[s], s, \ldots), \;s:=s\modnop{+}1, \;w:=w+1]$
\i $(M_1\neq \varepsilon)\wedge (seq(\hat M_1)=r)\;[Out\,!\,info(\hat M_1),\; 
r:=r\modnop{+}1,\; M_1:=M_1']$
\i $(M_2\neq \varepsilon)\wedge(ack (\hat M_2) \in [b, s[
)\;[b:=ack (\hat M_2)\modnop{+}1,\; 
w:=s\modnop{-}b,\; M_2:=M_2']$,
where $\forall\,i,j\in \{0,n-1\}\;\;i\modnop{-}j\eam
     i{-}j$, if $i{-}j \in  \{0,n-1\}$, and  
$n+i{-}j$, otherwise
\i $[M_1:=
   M_1\cdot \varphi(x[b], b,\ldots)
  ,\ldots\;,
   M_1:=M_1\cdot \varphi(x[s\modnop{-}1], s\modnop{-}1, 
   \ldots)]$
\i $(M_1\neq \varepsilon)\;[M_1:=M'_1]$
\i $(M_2\neq \varepsilon)\;[M_2:=M'_2]$
\i $[M_2:=
   M_2\cdot \varphi(\ldots, \ldots, r\modnop{-}1)]$
\ei
where
dots denote unessential components of expressions, and the symbols $M_i$,
$\hat M_i$, $M'_i$, $\cdot$ and $\varepsilon$
have the following sense:
\bi
\i $M_1$ and $M_2$ are variables of the process $Channel$, and values of these variables are lists of frames which were received by the process 
$Channel$ 
($M_i$
holds frames received from $Agent_i$), 
every received frame is added to the end
of a corresponded list
\i $\hat M_i\;(i=1,2)$
is an expression, a value of which is equal to the first element
of the list $M_i$
\i $M'_i\;(i=1,2)$
 is an expression, a value of which is equal to the list 
 $M_i$ without
its first element
\i $\cdot$ is a function of an addition of a frame to the end of a list
\i $\varepsilon$ 
is a constant, a value of which is an empty list.
\ei

For a proof that the process $P$ is observationally equivalent to the process $B_{n-1}$, 
we define a formula $b_{s_1s_2}$
where $s_1$ is a unique state of $P$
and $s_2$ is a unique state of $B_{n-1}$
as a conjunction of the following formulas:
\bi
\i $(M_1\neq \varepsilon)\wedge (seq(\hat M)=r)
    \;\Rightarrow\;u>0$
\i $\forall\; f\in M_1\;
   info(f)=x[seq(f)]$
   \i $\forall\; f\in M_2\quad ack(f)\in [b\modnop{-}1,r[$
\i $[r,s[\;\subseteq \;[b,s[$
\i $w=s\modnop{-}b\leq n-1$
\i $u=s\modnop{-}r\leq w$
\i  if a value of $M_2$ is $f_1\cdot\ldots\cdot f_k$, 
then the sequence $ack(f_1)\quad\ldots\quad ack(f_k)$
is monotonically increasing (mod $n$) subsequence of $[b\modnop{-}1,r[$
\ei
(the last record is not a formula, but can be represented
by a formula, we omit this representation).

It is not so diffcult to check that $b_{s_1s_2}$ satisfies the conditions of theorem \ref{th1}
and this proves that 
the process $P$ is observationally equivalent to $B_{n-1}$.

 \section{Conclusion}

The concept of a process with message passing which is presented in this paper
can be considered as a formal model of a communicating program without recursion. 
In the paper we have established suffcient conditions of observational
equivalence of processes. The next steps of investigations in this area can be the
following:
find necessary and suffcient conditions of observational equivalence of message passing
processes,
generalize the proposed concept of a process with message passing for formal
modeling of communicating programs with recursion, and find necessary and
suffcient conditions of observational equivalence of such processes.

\newpage

\section*{Appendix}

\section{A description of 
sliding window protocol}

\subsection{Frames}

Each frame $f$, which is sent by any of the agents, contains a packet $x$, and a couple of numbers:
\bi
\i a number $s \in {\bf Z}_n\eam\{0,1,\ldots, n-1\}$
   (where $n$ is a fixed integer), which is
associated with the packet $x$ and with the frame $f$, and
\i a number $r \in {\bf Z}_n$, 
   which is a number associated with a last received undistorted frame.
   \ei
To build a frame, a function $\varphi$ is used, i.e. a frame has the form $\varphi(x, s, r)$.

To extract the components $x$, $s$, $r$ 
from the frame $\varphi(x, s, r)$, the functions
$info$, $seq$ and $ack$
are used, these functions have the following properties:
$$info(\varphi(x, s, r)) = x,
\quad
 seq(\varphi(x, s, r)) = s,
 \quad 
 ack(\varphi(x, s, r)) = r$$
\subsection{Window}

The set of variables of 
an agent contains an array $x[n]$. 
Values of some
components of this array are
 packets which are sent, but not yet acknowledged. 
 A set of components
of the array $x$, which contain such packets at a current time, is called a {\bf window}.

Three variables of the agent are related to the window:
$b$ (a lower bound of the window),
$s$ (an upper bound of the window), and
$w$ (a number of packets in the window).
Values of these variables belong to the set ${\bf Z}_n$. At the initial moment  values of $b$, $s$
and $w$ are equal to 0. 
At any moment the window can be empty (if $b = s$), or
not empty (if $b \neq s$). 
In the last case the window consists of elements of $x$ with
indices from the set $[b, s[$, where $[b, s[$ 
denotes the set
\bi
\i $\{b, b + 1, \ldots, s-1\}$, if $b < s$, and
\i $\{b, b + 1, \ldots, n\} \cup \{0, 1, \ldots, s -1\}$,
   if $s < b$.
\ei

Adding a new packet to the window is performed by an 
execution of the following
actions: this packet is written in the component $x[s]$, 
$s$ is increased by 1 modulo
$n$ (i.e. a new value of $s$ is assumed to be $s+1$, 
if $s < n-1$, and 0, if $s = n-1$),
and $w$ is increased by 1. Removing a packet from the window is performed by
an execution of the following operations: $b$ is increased by 1 modulo $n$, and $w$ is
decreased by 1 (i.e. it is removed a packet whose number is equal to the lower
bound of the window).

If an agent 
received a frame, the third component $r$ 
of which (i.e. a
number of an acknowledgment) is such that $r \in [b, s[$, then all packets in the
window with numbers from $[b, r[$ are considered as acknowledged and are removed
from the window (even if their 
acknowledgments were not received).

\subsection{Flow graph}

A relation between subprocesses 
of sliding window protocol
is represented by the flow graph:
$$\by 
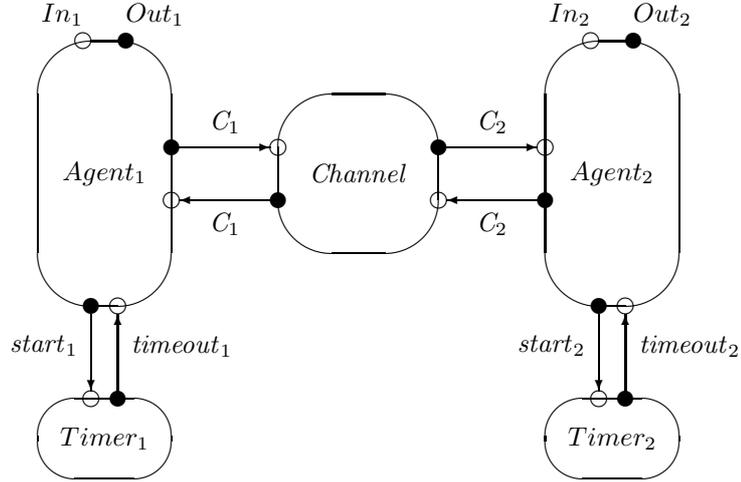
\begin{picture}(0,120)


\put(-30,60){\circle{6}}
\put(30,60){\circle*{6}}
\put(-30,40){\circle*{6}}
\put(30,40){\circle{6}}
\put(-70,60){\circle*{6}}
\put(70,60){\circle{6}}
\put(-70,40){\circle{6}}
\put(70,40){\circle*{6}}

\put(-103,100){\circle{6}}
\put(-87,100){\circle*{6}}

\put(87,100){\circle{6}}
\put(103,100){\circle*{6}}

\put(-100,-35){\circle{6}}
\put(-90,-35){\circle*{6}}

\put(-100,0){\circle*{6}}
\put(-90,0){\circle{6}}

\put(-100,-3){\vector(0,-1){29}}
\put(-90,-32){\vector(0,1){29}}

\put(-105,-15){\makebox(0,0)[r]{${\it start}_1$}}
\put(-85,-15){\makebox(0,0)[l]{${\it timeout}_1$}}

\put(90,-35){\circle{6}}
\put(100,-35){\circle*{6}}

\put(90,0){\circle*{6}}
\put(100,0){\circle{6}}

\put(90,-3){\vector(0,-1){29}}
\put(100,-32){\vector(0,1){29}}

\put(85,-15){\makebox(0,0)[r]{${\it start}_2$}}
\put(105,-15){\makebox(0,0)[l]{${\it timeout}_2$}}

\put(-50,65){\makebox(0,0)[b]{${\it C}_1$}}
\put(-50,35){\makebox(0,0)[t]{${\it C}_1$}}

\put(50,65){\makebox(0,0)[b]{${\it C}_2$}}
\put(50,35){\makebox(0,0)[t]{${\it C}_2$}}

\put(0,50){\oval(60,60)}
\put(0,50){\makebox(0,0){${\it Channel}$}}

\put(-95,50){\oval(50,100)}
\put(-95,50){\makebox(0,0){$Agent_1$}}

\put(-95,-50){\oval(50,30)}
\put(-95,-50){\makebox(0,0){$Timer_1$}}

\put(95,-50){\oval(50,30)}
\put(95,-50){\makebox(0,0){$Timer_2$}}

\put(95,50){\oval(50,100)}
\put(95,50){\makebox(0,0){$Agent_2$}}

\put(-67,60){\vector(1,0){34}}
\put(-33,40){\vector(-1,0){34}}
\put(33,60){\vector(1,0){34}}
\put(67,40){\vector(-1,0){34}}
\put(-103,110){\makebox(0,0)[r]{$In_1$}}
\put(-87,110){\makebox(0,0)[l]{$Out_1$}}

\put(87,110){\makebox(0,0)[r]{$In_2$}}
\put(103,110){\makebox(0,0)[l]{$Out_2$}}

\end{picture}
\\
\vspace{14mm}\ey
$$
$$\;$$

\subsection{Timers}

Each component $x[i]$ of the array $x$ is associated with a timer, which determines
a duration of waiting of an acknowledgement from another agent of a receiving
of the packet contained in the component $x[i]$. The combination of these timers
is considered as a process $Timer$, which has an array $t [n]$ of boolean variables.
The process $Timer$ has one state and 
transitions which are labeled by the following 
operators:
\bi
\i $[start ? i, t\,[i]:=1]$
\i $[stop?i,\; t\,[i]:=0]$
\i $(t\,[j]=1)[ timeout\,!\,j, t\,[j]:=0]$
(where $j = 0,\ldots, n-1$)
\ei
An initial condition is $t = (0,\ldots,0)$.

If an agent has received an object with a name $timeout$ from a timer, then
the agent sends again all packets from its window.

\subsection{Agents}

A behavior of each agent is described by the same process, combining functions
of a sender and a receiver. This behavior can be represented by the following
flowchart.
$$\by
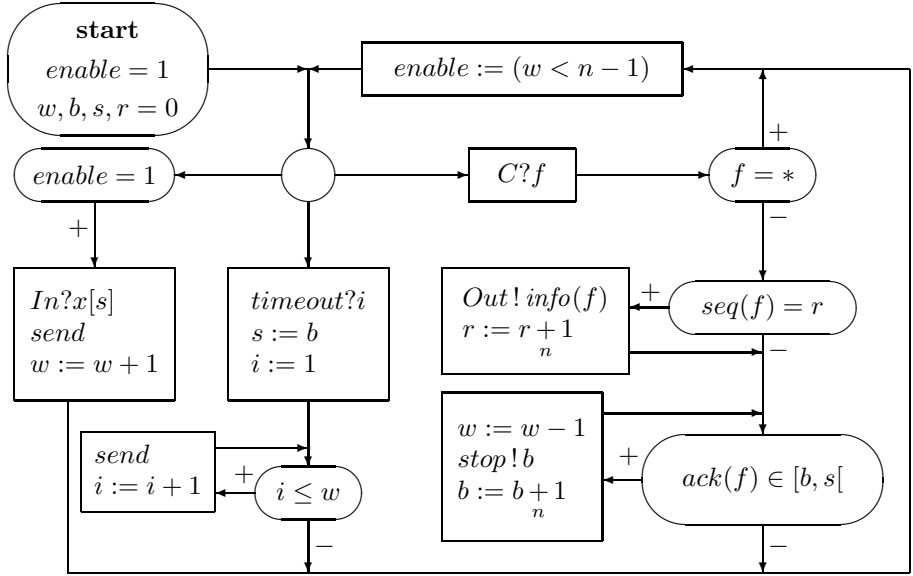
\begin{picture}(110,190)

\put(-75,160){\oval(75,50)}
\put(-75,175){\makebox(0,0){${\bf start}$}}
\put(-75,160){\makebox(0,0)[c]{$enable=1$}}
\put(-75,145){\makebox(0,0)[c]{$w,b,s,r=0$}}

\put(0,60){\pu{30}{25}}
\put(0,60){\makebox(0,0){$\by timeout?i\\
s:=b\\i:=1\ey $}}

\put(-80,120){\oval(60,20)}
\put(-80,120){\makebox(0,0){$enable=1$}}

\put(-80,60){\pu{30}{25}}
\put(-80,60){\makebox(0,0){$\by In?x[s]
\\send\\
w:=w+1\ey $}}

\put(-60,8){\pu{25}{15}}
\put(-60,8){\makebox(0,0){$\by send\\ i:=i+1
\ey $}}

\put(0,0){\oval(40,20)}
\put(0,0){\makebox(0,0){$i\leq w$}}

\put(0,120){\oval(20,20)}

\put(85,65){\pu{35}{20}}
\put(85,65){\makebox(0,0){$\by Out\,!\, {\it info} (f)\\
r:=r\modnop{+}1
\ey $}}

\put(80,10){\pu{30}{28}}
\put(80,10){\makebox(0,0){$\by w:=w-1

\\stop\,!\,b\\b:=b\modnop{+}1
\ey $}}

\put(170,5){\oval(90,34)}
\put(170,5){\makebox(0,0){$\by ack (f) \in [b, s[
\ey $}}

\put(170,70){\oval(70,20)}
\put(170,70){\makebox(0,0){$
 seq (f)=r
$}}

\put(170,120){\oval(40,20)}
\put(170,120){\makebox(0,0){$f=*
$}}

\put(80,120){\pu{20}{10}}
\put(80,120){\makebox(0,0){$C?f
$}}

\put(80,160){\pu{60}{10}}
\put(80,160){\makebox(0,0){$enable:=(w<n-1)
$}}

\put(-37.5,160){\vector(1,0){37.5}}

\put(0,160){\vector(0,-1){30}}
\put(0,110){\vector(0,-1){25}}
\put(0,35){\vector(0,-1){25}}
\put(0,-10){\vector(0,-1){20}}

\put(-90,-30){\line(1,0){315}}
\put(-90,-30){\line(0,1){65}}
\put(225,-30){\line(0,1){190}}
\put(225,160){\vector(-1,0){85}}
\put(20,160){\vector(-1,0){20}}
\put(-10,120){\vector(-1,0){40}}
\put(10,120){\vector(1,0){50}}
\put(100,120){\vector(1,0){50}}

\put(170,130){\vector(0,1){30}}
\put(170,110){\vector(0,-1){30}}
\put(-80,110){\vector(0,-1){25}}
\put(170,60){\vector(0,-1){38}}
\put(170,-12){\vector(0,-1){18}}

\put(125,5){\vector(-1,0){15}}

\put(110,30){\vector(1,0){60}}

\put(135,70){\vector(-1,0){15}}
\put(120,53){\vector(1,0){50}}

\put(-35,17){\vector(1,0){35}}

\put(-20,0){\vector(-1,0){15}}

\put(176,136){\makebox(0,0){$+$}}
\put(176,104){\makebox(0,0){$-$}}
\put(176,54){\makebox(0,0){$-$}}
\put(176,-20){\makebox(0,0){$-$}}
\put(6,-20){\makebox(0,0){$-$}}
\put(-86,100){\makebox(0,0){$+$}}
\put(-24,7){\makebox(0,0){$+$}}
\put(120,14){\makebox(0,0){$+$}}
\put(128,76){\makebox(0,0){$+$}}

\end{picture}\\
\vspace{10mm}
\ey
$$ 

where
\bi
\i $send$ is an abbreviation of the list of AOs
   $\c{C\,! \,\varphi (x[s], s, 
   r\modnop{-}1)\\start\;!\;s\\
   s:=s\modnop{+}1}$
\i $*$ is a special notation for a distorted message, and
\i a value of the variable $enable$ is 1, if the agent can receive a new packet from
his network level (i.e. $w < n - 1$), and 0, otherwise.
\ei

Processes $Agent_1$ and $Agent_2$ 
are obtained by a simple transformation of this
flowchart, 
and by an addition of corresponded index 
(1 or 2) to its variables and names.

\subsection{A proof of theorem 1}

Since $X_{P_1}\cap X_{P_2}=\emptyset$,
then there is a natural bijection between
$X_{P_1}^\bullet \times X_{P_2}^\bullet$ 
and $(X_{P_1}\cup X_{P_2})^\bullet$.
Below we identify these two sets.

We define the relation
$\mu\subseteq S_{P_1}^r \times S_{P_2}^r$
as follows:
$$\mu\eam
\{(\xi_1, \xi_2) \in X_{P_1}^\bullet \times X_{P_2}^\bullet
\mid 
b_{at_{P_1}^{\xi_1}at_{P_2}^{\xi_2}}^{(\xi_1,\xi_2)}=1
\}\cup\{(P_1^0, P_2^0)\}.$$

We prove that $\mu$ satisfies the conditions 
from section \ref{definition}.

\bn
\i The condition $(P_1^0, P_2^0)\in \mu$ 
follows from the definition of $\mu$.
\i Let $(v_1, v_2)\in \mu$ and $v_1\ra{\tau}v'_1$.
We must prove that
\be{fdgd4444}
\exists\,v'_2: v_2\ra{\tau^*}v'_2,\;
(v'_1, v'_2)\in \mu
\ee

We consider separately the cases
$v_1=P^0_1$ and $v_1\neq P^0_1$.

If $v_1=P^0_1$, then $v_2=P^0_2$, and
according to definition of the graph
 $P_1^r$
(section \ref{defpr}), 
$\exists\,\xi_1\in X_{P_1}^\bullet:
\l{P_1}^{\xi_1}=1$ and the graph $P_1^r$ 
has the edge
$\xi_1\ra{\tau}\xi'_1=v'_1$,
i.e.
$(\xi_1, \tau, \xi'_1)$ is a realization of a 
transition 
$s_{P_1}^0\ra{O_1}s'_1$
from $T_{P_1}$, where  $O_1$ is an 
internal operator. 

According to item \ref{sdfsa33}
in the theorem, there exists a set
$\{s_{P_2}^0\ra{T_i}s_2^i\mid i\in\Im\}$
of CTs of process   
   $P_2$, 
such that
\be{fdgdsfgfdsgr4r}
b_{s_{P_1}^0s_{P_2}^0} \wedge \; \l{O_1}
\leq
\bigvee\limits_{i\in\Im}
\deyc{(O_1, O_{T_i})}
{b_{s'_1s^{i}_2}}
\ee

Since  $\l{P_2}\neq \bot$, then
$\exists\,\xi_2\in X_{P_2}^\bullet$:
$\l{P_2}^{\xi_2} = 1$, so
\be{fdgdsfgfdsgr4r1}
1=\l{P_1}^{\xi_1} \wedge\l{P_2}^{\xi_2}=
(\l{P_1}\wedge \l{P_2})^{(\xi_1,\xi_2)}\leq
b_{s^0_{P_1}s^0_{P_2}}^{(\xi_1,\xi_2)}\ee
(the last inequality holds
according to property \ref{sdfsa332} 
in the statement of the theorem).

According to the definition of a realization of a
transition, the equality
$\l{O_1}^{\xi_1}=1$ holds.
This equality,
\re{fdgdsfgfdsgr4r} and \re{fdgdsfgfdsgr4r1},
imply that there is  $i\in \Im$ such that
\be{s44dfsdfsd}\Big(\deyc{(O_1, O_{T_i})}
{b_{s'_1s^{i}_2}}\Big)^{(\xi_1,\xi_2)} = 1\ee

It is easy to prove that the equality
\be{sdfsd55}\Big(\deyc{(O_1, O_{T_i})}
{b_{s'_1s^{i}_2}}\Big)^{(\xi_1,\xi_2)} = 
{b_{s'_1s^{i}_2}}^{(\xi_1\cdot O_1,\xi_2\cdot
O_{T_i})}
\ee
holds. This equality is an analogue of the equality 
in the end of section \ref{concatt}, 
and is proved by induction on the total 
number of AOs in $[O_1]$ and $[O_2]$.

\re{s44dfsdfsd} and \re{sdfsd55} imply that
\be{sdfsd5345}
{b_{s'_1s^{i}_2}}^{(\xi_1\cdot O_1,\xi_2\cdot
O_{T_i})}=1
\ee

By the definition of $\mu$ and $\xi_2$, 
the statement \re{fdgd4444} in this case
($v_1=P^0_1$)
follows from the statement
\be{dfgfdgdrr}\exists\,\xi'_2: \xi_2 \ra{\tau^*}\xi'_2,\;
b_{at_{P_1}^{\xi'_1}at_{P_2}^{\xi'_2}}^{(\xi'_1,\xi'_2)}=1\ee

Define 
$\xi'_2\eam (\xi_2\cdot (at_{P_2}:=
s_2^i))\cdot O_{T_i}$. Since $at_{P_1}^{\xi'_1}=
s'_1$, and
$\xi'_1=(\xi_1\cdot(at_{P_1}:=s'_1))\cdot O_1$,
then \re{dfgfdgdrr} follows from the statements
\be{dfgfd2gdrr}
\xi_2 \ra{\tau^*}(\xi_2\cdot (at_{P_2}:=
s_2^i))\cdot O_{T_i}\ee
\be{dfgfd2gdr3r}
b_{s'_1s^i_2}^{((\xi_1\cdot(at_{P_1}:=s'_1))\cdot O_1,(\xi_2\cdot (at_{P_2}:=
s_2^i))\cdot O_{T_i})}=1\ee

\re{dfgfd2gdrr} follows from 
the definitions of concepts 
of a CT and a concatenation 
of operators and from the statements
$at_{P_2}^{\xi_2}=s_{P_2}^0$
and $\l{O_{T_i}}^{\xi_2}=1$.
The first of these statements 
follows from the equality
$\l{P_2}^{\xi_2}=1$,
and the second is justified as follows.
The definition of formulas of the form
$(O_1,O_2)\cdot b$
implies that the statement
\re{s44dfsdfsd}
can be rewritten as
\be{sfdgfdsg4455}\Big(\l{O_1}\wedge \l{O_{T_i}}
\wedge b \Big)^{(\xi_1,\xi_2)} = 1\ee
where $b$ is some formula.
Since $X_{P_1}\cap X_{P_2}=\emptyset$,
then \re{sfdgfdsg4455} implies the desired statement
$\l{O_{T_i}}^{\xi_2}=1$.

\re{dfgfd2gdr3r} follows from 
\re{sdfsd5345} and from the assumption that
 $at_{P_1}$ and $at_{P_2}$
do not occur in $b_{s'_1s^i_2}$, 
$O_1$ and $O_{T_i}$.

Thus, in the case
$v_1 = P_1^0$ the property
\re{fdgd4444}  holds.

In the case $v_1\neq P_1^0$ the property
\re{fdgd4444} can be proved similarly.

\i Let $(v_1, v_2)\in \mu$ and $v_1\ra{a}v'_1$,
where $a\neq \tau$. We must prove that
\be{1fdgd4444}
\exists\,v'_2: v_2\ral{\tau^*a\tau^*}v'_2,\;
(v'_1, v'_2)\in \mu\ee

\bn

\i At first consider the case
$v_1=P^0_1$ and è $a=\alpha?d$.

If $v_1=P^0_1$, then $v_2=P^0_2$, and
according to the definition of the graph $P_1^r$
(section \ref{defpr}), 
$\exists\,\xi_1\in X_{P_1}^\bullet:
\l{P_1}^{\xi_1}=1$ and the graph $P_1^r$ 
has the edge
$\xi_1\ra{a}\xi'_1=v'_1$,
i.e.
$(\xi_1, a, \xi'_1)$ is a realization of a transition 
$t$ of the form
$s_{P_1}^0\ra{O_1}s'_1$
from $T_{P_1}$, where $O_1$ is an input operator. 
Using the notation introduced at the end of section
\ref{realper}, we can write
$\xi'_1 = \xi_1\cdot t^d$.

Just as in the preceding item, we prove that
$\exists\,\xi_2\in X_{P_2}^\bullet$:
$\l{P_2}^{\xi_2} = 1$,
and there exists a CT $s_{P_2}^0\ra{T_i}s^i_2$
of the process $P_2$, 
such that the equality
\be{1s44dfsdfsd}\Big(\deyc{(O_1, O_{T_i})}
{b_{s'_1s^{i}_2}}\Big)^{(\xi_1,\xi_2)} = 1\ee
holds, which should be understood in the following sense: for each of valuation
$\xi\in (X_{P_1}\cup X_{P_2}\cup
\{z\})^\bullet$ (where $z$ is a variable,
referred in the item \ref{peremz} 
of the definition from section \ref{theor}, 
we can assume that
$z\not\in ((X_{P_1}\cup X_{P_2})$),
coinciding with 
$\xi_i$ on $X_{P_i}\;\;(i=1,2)$,
the equality 
 $\Big(\deyc{(O_1, O_{T_i})}
{b_{s'_1s^{i}_2}}\Big)^{\xi} = 1$ holds.
In particular, 
\re{1s44dfsdfsd} implies that
$O_{T_i}$ is an input operator,
and 
$N_{O_{T_i}} = N_{O_1}=\alpha$.

Define $\xi'_2\eam \xi_2\cdot T_i^d$.
It is easy to prove that
$\xi_2 \ra{\tau^*a \tau^*}\xi'_2$, and
the statement
\re{1fdgd4444} in the case
$v_1=P^0_1$
follows from the equality
\be{1dfgf33dgdrr}
b_{s'_1s^i_2}^{(\xi_1\cdot t^d,\xi_2\cdot T_i^d)}=1\ee which is justified as follows.

In this case
$O_1$ and $O_{T_i}$ can be represented as concatenation of the form
$$O_1=(O'_1\cdot[\alpha?x])\cdot O''_1,\quad
O_{T_i} = (O'_{T_i}\cdot[\alpha?y])\cdot  
O''_{T_i}$$ 
Definition of formulas of the form
\re{fgvfdsgdsfgdsfgd}
implies that 
\be{1s44dfsdfsd11}
\by
\deyc{(O_1, O_{T_i})}{b_{s'_1s^{i}_2}}=\\=
\deyc{\Big((O'_1\cdot[\alpha?x])\cdot O''_1, 
(O'_{T_i}\cdot[\alpha?y])\cdot  
O''_{T_i}\Big)}{b_{s'_1s^{i}_2}}=\\
=
\deyc{\Big(O'_1\cdot[\alpha?x], 
O'_{T_i}\cdot[\alpha?y]  
\Big)}{\Big((O''_1,O''_{T_i})\cdot b_{s'_1s^{i}_2}\Big)}
=\\
=
\deyc{(O'_1, 
O'_{T_i} 
)}{\Big(\big((O''_1,O''_{T_i})\cdot b_{s'_1s^{i}_2}\big)
(z/x,z/y)\Big)}\ey\ee

\re{1s44dfsdfsd} and \re{1s44dfsdfsd11}
imply the equality
$$\Big(\big((O''_1,O''_{T_i})\cdot b_{s'_1s^{i}_2}\big)
(z/x,z/y)\Big)^{(\xi_1\cdot O'_1,\xi_2\cdot O'_{T_i})} = 1$$
Its special case is the equality
$$\Big(\big((O''_1,O''_{T_i})\cdot b_{s'_1s^{i}_2}\big)
(d/x,d/y)\Big)^{(\xi_1\cdot O'_1,\xi_2\cdot O'_{T_i})} = 1$$
The last equality can be rewritten as
$$\Big((O''_1, O''_{T_i})\cdot b_{s'_1s^{i}_2}
\Big)^{(\xi_1\cdot O'_1\cdot (x:=d),\xi_2\cdot O'_{T_i}
\cdot (y:=d))} = 1$$
whence it follows that
\be{sfdgdsfsdfsdadfsdsf}\Big(b_{s'_1s^{i}_2}
\Big)^{(\xi_1\cdot O'_1\cdot (x:=d)\cdot O''_1,\xi_2\cdot O'_{T_i}
\cdot (y:=d)\cdot O''_{T_i})} = 1\ee

It is easy to see that the left side 
of \re{sfdgdsfsdfsdadfsdsf} coincides with 
the left side of the equality \re{1dfgf33dgdrr}.

Thus, in the case
$v_1 = P_1^0$ and $a=\alpha?d$
the property
\re{1fdgd4444}   is proven.

In the case $v_1\neq P_1^0$ and $a=\alpha?d$
the property
\re{1fdgd4444} can be proved similarly.

\i Now we prove \re{1fdgd4444}, 
when $a=\alpha!d$.
As in the previous item, 
we consider only the case
$v_1=P^0_1$.

If $v_1=P^0_1$, then $v_2=P^0_2$, and
\bi
\i $\exists\,\xi_1\in X_{P_1}^\bullet:
\l{P_1}^{\xi_1}=1$ and the graph  $P_1^r$ 
has the edge
$\xi_1\ra{a}\xi'_1=v'_1$,
i.e.
$(\xi_1, a, \xi'_1)$ is a realization of a transition
$t\in T_{P_1}$ of the form
$s_{P_1}^0\ra{O_1}s'_1$,
where $O_1$ is an output operator
\i $\exists\,\xi_2\in X_{P_2}^\bullet$:
$\l{P_2}^{\xi_2} = 1$,
and there exists a CT $s_{P_2}^0\ra{T_i}s^i_2$
of the process $P_2$, such that 
\be{11s44dfsdfsd}\Big(\deyc{(O_1, O_{T_i})}
{b_{s'_1s^{i}_2}}\Big)^{(\xi_1,\xi_2)} = 1\ee
\ei

\re{11s44dfsdfsd} implies that
$O_{T_i}$ is an output operator,
and $N_{O_{T_i}} = N_{O_1}=\alpha$.

Define  $\xi'_2\eam \xi_2\cdot T_i$.
For a proof of \re{1fdgd4444} 
it is enough to prove the statements
\be{sdfsdfsdfsfs}
\xi_2 \ra{\tau^*a \tau^*}\xi'_2
\ee
\be{11dfgf33dgdrr}
b_{s'_1s^i_2}^{(\xi_1\cdot t,\xi_2\cdot T_i)}=1\ee 

In this case
$O_1$ and $O_{T_i}$ 
can be represented as concatenations of the form
\be{sdfsdafsad}
O_1=(O'_1\cdot[\alpha!e_1])\cdot O''_1\ee
\be{sdfsdafsad1}
O_{T_i} = (O'_{T_i}\cdot[\alpha!e_2])\cdot  
O''_{T_i}\ee

The definition of formulas of the form
\re{fgvfdsgdsfgdsfgd}
implies that
\be{21s44dfsdfsd11}
\by
\deyc{(O_1, O_{T_i})}{b_{s'_1s^{i}_2}}=\\=
\deyc{\Big((O'_1\cdot[\alpha!e_1])\cdot O''_1, 
(O'_{T_i}\cdot[\alpha!e_2])\cdot  
O''_{T_i}\Big)}{b_{s'_1s^{i}_2}}=\\
=
\deyc{\Big(O'_1\cdot[\alpha!e_1], 
O'_{T_i}\cdot[\alpha!e_2]  
\Big)}{\Big((O''_1, O''_{T_i})\cdot b_{s'_1s^{i}_2}\Big)}
=\\
=
\deyc{(O'_1, 
O'_{T_i} 
)}{\c{e_1=e_2\\(O''_1,O''_{T_i})\cdot b_{s'_1s^{i}_2}}
}\ey\ee

\re{11s44dfsdfsd} and \re{21s44dfsdfsd11}
imply the equality
$${\c{e_1=e_2\\(O''_1,O''_{T_i})\cdot b_{s'_1s^{i}_2}}}^{(\xi_1\cdot O'_1,\xi_2\cdot O'_{T_i})}
= 1
$$
from which it follows that
\be{dfgfdsgdsf5554}
e_1^{\xi_1\cdot O'_1}=e_2^{\xi_2\cdot O'_{T_i}}
\ee
\be{dfgfdsgdsf55541}
\Big((O''_1, O''_{T_i})\cdot b_{s'_1s^{i}_2}\Big)^{(\xi_1\cdot O'_1,\xi_2\cdot O'_{T_i})}
= 1
\ee

By assumption,
$(\xi_1, \alpha!d, \xi'_1)$ is a realization of the transition
$s_{P_1}^0\ra{O_1}s'_1$.
From the representation of 
$O_1$  as a concatenation
\re{sdfsdafsad} it follows that
$d=e_1^{\xi_1\cdot O'_1}$, 
whence, according to \re{dfgfdsgdsf5554} we get
the equality
$d=e_2^{\xi_2\cdot O'_{T_i}}$.
From this and from a representation of 
$O_{T_i}$ as a concatenation 
\re{sdfsdafsad1} 
it follows that 
$(\xi_2, \alpha!d, \xi_2\cdot T_i)$ 
is a realization of the CT 
$T_i$. 
Since $\xi_2\cdot T_i = \xi'_2$ and 
$\alpha!d=a$, 
then it follows that we are justified the
statement \re{sdfsdfsdfsfs}.

The statement  \re{11dfgf33dgdrr} follows from 
\re{dfgfdsgdsf55541}.

Thus, in the case
$v_1 = P_1^0$ and $a=\alpha!d$
the property
\re{1fdgd4444}   is proven.

In the case 
$v_1\neq P_1^0$ and $a=\alpha!d$
the property
\re{1fdgd4444} can be proved similarly

\en

\en

The symmetrical conditions on the relation 
 $\mu$ (i.e.,  second parts of the conditions on $\mu$, 
presented in second and third items in section 
 \ref{definition}) can be proved similarly. $\blackbox$

 \section{An example of a process defined
with use of parallel composition and
restriction}

In this subsection we describe a process 
which is defined
with use of the operations of 
parallel composition and
restriction.
This process is an implementation 
of a distributed algorithm 
of separation of sets.
The problem of separation of sets has the 
following form.
Let $U,V$ be a pair of finite disjoint sets, 
with each element $x \in U \cup V$ is 
associated with a number $weight(x)$, 
called a {\bf weight} of this element. 
It is need to convert this pair 
to a pair of sets $U', V'$, such that
\bi
\i $|U|=|U'|,\quad
   |V|=|V'|$\\  (for each finite set $M$ 
the notation $|M|$ denotes a number 
of elements in $M$) 
\i $\forall\,u \in U', \forall\,v \in V'\quad
weight(u) \leq weight(v)$.
\ei 

Below we shall call $U$ and $V$
as the left set and the right set, respectively.

The problem of separation of sets 
can be solved by an execution of 
several sessions of exchange elements 
between these sets. 
Each session consists of the following actions: 
\bi 
\i find an element $mx$ with 
   a maximum weight in the left set 
\i find an element $mn$ with minimum weight 
   in the right set 
\i transfer 
   \bi 
   \i $mx$ from the left set to the right set, and 
   \i $mn$ from the right set to the left set. 
   \ei 
\ei 

To implement this idea it is proposed 
a distributed algorithm, defined 
as a process of the form 
\be{rrwarefgtee333}(Small\mid Large)
    \setminus  \{\alpha,\beta\}\ee
where 
\bi 
\i the process $Small$ executes 
   operations associated with 
   the left set, and 
\i the process $Large$ executes
   operations associated with 
   the right set. 
\ei 

The restriction of the actions with names 
$\alpha$ and $\beta$ in \re{rrwarefgtee333}
means that a transmission of objects with 
names $\alpha$ and $\beta$ can be executed
only between the subprocesses $Small$ 
and $Large$, i.e. such objects can not be 
transmitted outside the process 
\re{rrwarefgtee333}.

A flow graph
(i.e. a relation between components)
corresponded to this process
has the form 
{\small
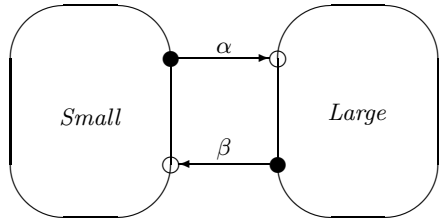
$$\by
\begin{picture}(0,100)

\put(50,50){\oval(60,80)}
\put(-50,50){\oval(60,80)}
\put(20,70){\circle{6}}
\put(20,30){\circle*{6}}
\put(-20,70){\circle*{6}}
\put(-20,30){\circle{6}}

\put(-17,70){\vector(1,0){34}}
\put(17,30){\vector(-1,0){34}}

\put(0,72){\makebox(0,0)[b]{$\alpha$}}
\put(0,32){\makebox(0,0)[b]{$\beta$}}

\put(-50,45){\makebox(0,0)[b]{${\it Small}$}}
\put(50,45){\makebox(0,0)[b]{${\it Large}$}}

\end{picture}
\ey
$$ } 

Below we shall use the following notations: 
for each subset $W\subseteq U\cup V$
   the records
   $\max(W)$ and $\min(W)$
   denote an element of $W$ 
   with maximum and minimum weight, respectively. 
A similar meaning have the records 
$\max(W)$ and $\min(W)$,
where
$W$ is a 
a variable whose values are 
subsets of $U \cup V$.

The process {\it Small} 
has the following form:

{\def\arraystretch{1}

{\small
\be{script343}\by
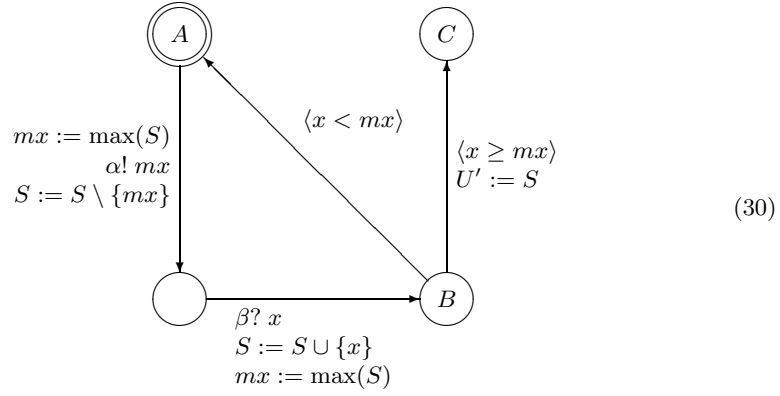
\begin{picture}(100,105)

\put(0,100){\oval(20,20)}
\put(0,100){\oval(24,24)}
\put(0,0){\oval(20,20)}
\put(100,100){\oval(20,20)}
\put(100,0){\oval(20,20)}

\put(0,100){\makebox(0,0){$A$}}
\put(100,0){\makebox(0,0){$B$}}
\put(100,100){\makebox(0,0){$C$}}

\put(0,88){\vector(0,-1){78}}
\put(10,0){\vector(1,0){80}}
\put(100,10){\vector(0,1){80}}
\put(93,7){\vector(-1,1){84}}

\put(-2,50){\makebox(0,0)[r]{$\begin{array}
{r}
mx:=\max(S)\\
\alpha!\;mx\\
S:=S \setminus \{mx\}
\ey$}}

\put(50,-2){\makebox(0,0)[t]{$\begin{array}{l}
\beta?\;x\\
S:=S\cup \{x\}\\
mx:=\max(S)
\ey$}}

\put(102,50){\makebox(0,0)[l]{$
\by \l{x \geq mx}\\ U':=S\ey
$}}

\put(65,67){\makebox(0,0){$\l{x < mx}$}}

\end{picture}
\\
\vspace{8mm}
\ey
\ee } 
}
(a double circle denotes an initial state).

An initial condition of the process $Small$
is $(S=U)$.

The process {\it Large}
has the following form:

{\def\arraystretch{1}

{\small

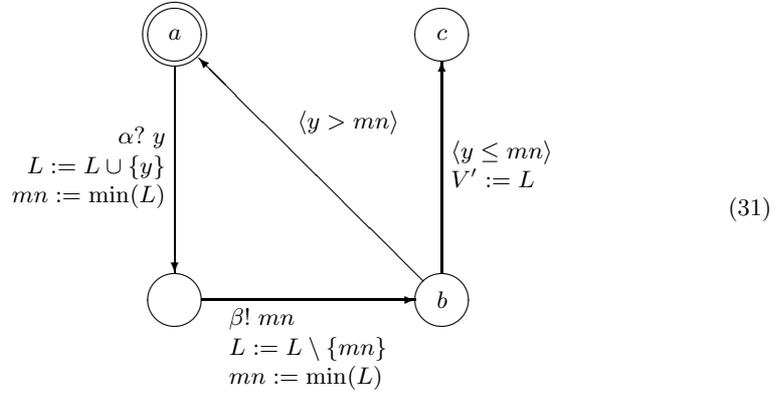
\be{large343}\by

\begin{picture}(100,105)

\put(0,100){\oval(20,20)}
\put(0,100){\oval(24,24)}
\put(0,0){\oval(20,20)}
\put(100,100){\oval(20,20)}
\put(100,0){\oval(20,20)}

\put(0,100){\makebox(0,0){$a$}}
\put(100,0){\makebox(0,0){$b$}}
\put(100,100){\makebox(0,0){$c$}}

\put(0,88){\vector(0,-1){78}}
\put(10,0){\vector(1,0){80}}
\put(100,10){\vector(0,1){80}}
\put(93,7){\vector(-1,1){84}}

\put(-2,50){\makebox(0,0)[r]{$\begin{array}{r}
\alpha?\;y\\
L:=L\cup \{y\}\\
mn:=\min(L)
\ey$}}

\put(50,-2){\makebox(0,0)[t]{$\begin{array}{l}
\beta!\;mn\\
L:=L\setminus \{mn\}\\
mn:=\min(L)
\ey$}}

\put(102,50){\makebox(0,0)[l]{$
\by \l{y\leq mn}\\V':=L\ey
$}}

\put(65,67){\makebox(0,0){$\l{y> mn}$}}

\end{picture}
\\
\vspace{8mm}
\ey
\ee
 } 
}

An initial condition of the process $Large$
is $(L=V)$.

A process 
which is obtained  by a simplification of 
the process 
\re{rrwarefgtee333}
has the following form:

{\def\arraystretch{1}
{\small
\be{gff54444777}\by
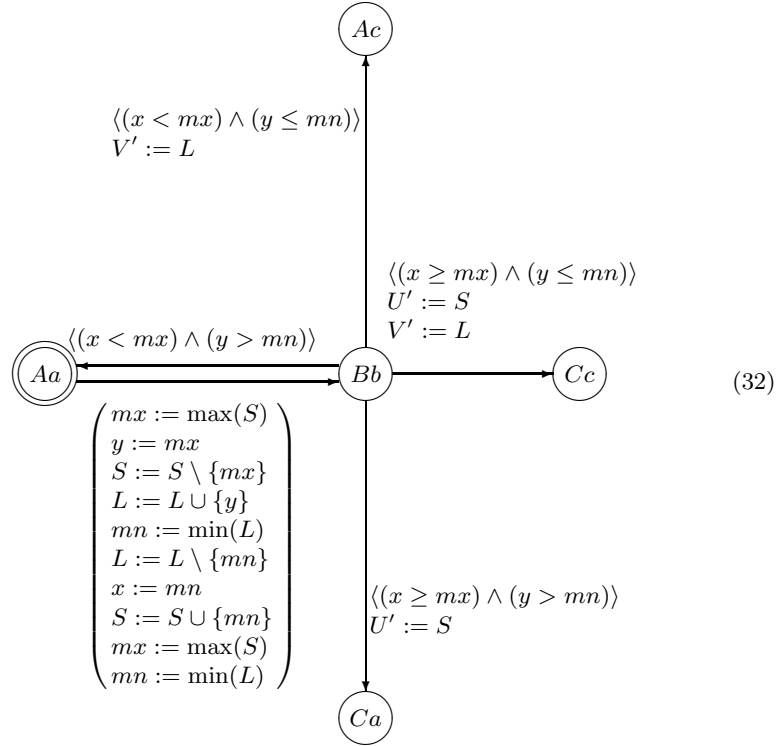
\begin{picture}(200,250)

\put(0,100){\oval(20,20)}
\put(0,100){\oval(24,24)}
\put(120,230){\oval(20,20)}
\put(120,100){\oval(20,20)}
\put(120,-30){\oval(20,20)}
\put(200,100){\oval(20,20)}

\put(0,100){\makebox(0,0){$Aa$}}
\put(120,-30){\makebox(0,0){$Ca$}}
\put(120,100){\makebox(0,0){$Bb$}}
\put(120,230){\makebox(0,0){$Ac$}}
\put(200,100){\makebox(0,0){$Cc$}}

\put(12,97){\vector(1,0){98}}
\put(110,103){\vector(-1,0){98}}

\put(120,110){\vector(0,1){110}}
\put(120,90){\vector(0,-1){110}}
\put(130,100){\vector(1,0){60}}

\put(55,90){\makebox(0,0)[t]{$\b{
mx:=\max(S)\\
y:=mx\\
S:=S\setminus \{mx\}\\
L:=L\cup\{y\}\\
mn:=\min(L)\\
L:=L\setminus \{mn\}\\
x:=mn\\
S:=S\cup\{mn\}\\
mx:=\max(S)\\
mn:=\min(L)
}$}}

\put(55,108){\makebox(0,0)[b]{$
\l{(x< mx)\wedge (y> mn)}
$}}

\put(175,110){\makebox(0,0)[b]{$\by 
\l{(x\geq mx) \wedge (y\leq mn)}
\\ U':=S \\ V':=L \ey$}}

\put(120,10){\makebox(0,0)[l]{$\by 
\l{(x\geq mx)\wedge (y> mn)}
\\U':=S\ey$}}

\put(120,190){\makebox(0,0)[r]{$\by\l{
(x<mx)\wedge (y\leq mn)
}\\V':=L\ey$}}

\end{picture}
\\
\vspace{15mm}
\ey
\ee
 } 
}

This simplified process allows 
to detect some simple flaws of the 
algorithm of separation of sets, for examle 
a possibility of a deadlock situation:
there are states of
the process \re{gff54444777}
(namely, $Ac$ and $Ca$)
such that
\bi 
\i there is no transitions starting 
   at these states 
   \i but falling into these states 
   is not a normal 
   completion of the process. 
\ei 

\section{Another example of
a simplification of a process} 
\label{umnozhenianadva}

Suppose we have a system ``multiplier'', 
which has 
\bi 
\i two input ports with the names 
   $In_1$ and $In_2$, and 
\i one output port with the name $Out$. 
\ei 

An execution of the multiplier is that it 
\bi 
\i receives on its input ports two values, and 
\i gives their product on the output port. 
\ei 
A behavior of the multiplier 
is described by the process $Mul$: 

{\def\arraystretch{1}
{\small
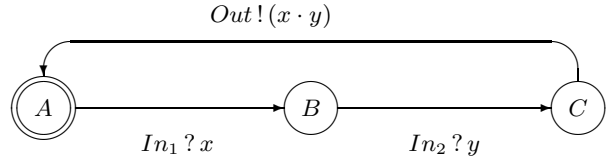
$$\by
\begin{picture}(210,40)

\put(0,0){\oval(20,20)}
\put(0,0){\oval(24,24)}
\put(0,0){\makebox(0,0)[c]{$A$}}

\put(100,0){\oval(20,20)}
\put(100,0){\makebox(0,0)[c]{$B$}}

\put(200,0){\oval(20,20)}
\put(200,0){\makebox(0,0)[c]{$C$}}

\put(12,0){\vector(1,0){78}}
\put(110,0){\vector(1,0){80}}

\put(50,-10){\makebox(0,0)[t]{$In_1\,?\,x$}}
\put(150,-10){\makebox(0,0)[t]{$In_2\,?\,y$}}

\put(0,15){\vector(0,-1){3}}
\put(200,15){\line(0,-1){5}}
\put(10,15){\oval(20,20)[tl]}
\put(190,15){\oval(20,20)[tr]}
\put(190,25){\line(-1,0){180}}
\put(85,30){\makebox(0,0)[b]{
$Out\,!\,(x\cdot y)$
}}

\end{picture}
\\
\vspace{0mm}
\ey
$$
 } 
}

Using this multiplier, we want to 
build a system ``calculator of a square'', 
whose behavior is described by 
the process $Square\_Spec$: 

{\small
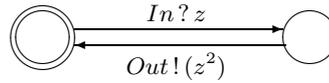
$$\by
\begin{picture}(100,20)

\put(0,0){\oval(20,20)}
\put(0,0){\oval(24,24)}
\put(100,0){\oval(20,20)}

\put(12,3){\vector(1,0){78}}
\put(91,-3){\vector(-1,0){79}}

\put(50,6){\makebox(0,0)[b]{$In\,?\,z$}}
\put(50,-6){\makebox(0,0)[t]{$Out\,!\,(z^2)$}}

\end{picture}
\\
\vspace{0mm}
\ey
$$ } 

The desired system is
a composition of
\bi 
\i the auxiliary system ``duplicator'' 
   having 
   \bi 
   \i an input port $In$, and 
   \i output ports $Out_1$ and $Out_2$ 
   \ei 
   behavior of which 
   is described by the process $Dup$: 

{\def\arraystretch{1}
{\small
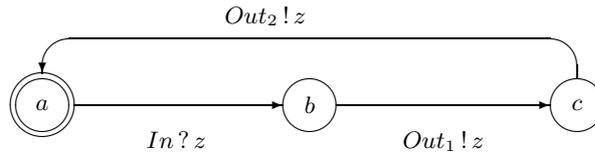
$$\by
\begin{picture}(210,40)

\put(0,0){\oval(20,20)}
\put(0,0){\oval(24,24)}
\put(0,0){\makebox(0,0)[c]{$a$}}

\put(100,0){\oval(20,20)}
\put(100,0){\makebox(0,0)[c]{$b$}}

\put(200,0){\oval(20,20)}
\put(200,0){\makebox(0,0)[c]{$c$}}

\put(12,0){\vector(1,0){78}}
\put(110,0){\vector(1,0){80}}

\put(50,-10){\makebox(0,0)[t]{$In\,?\,z$}}
\put(150,-10){\makebox(0,0)[t]{$Out_1\,!\,z$}}

\put(0,15){\vector(0,-1){3}}
\put(200,15){\line(0,-1){5}}
\put(10,15){\oval(20,20)[tl]}
\put(190,15){\oval(20,20)[tr]}
\put(190,25){\line(-1,0){180}}
\put(85,30){\makebox(0,0)[b]{$Out_2\,!\,z$
}}

\end{picture}
\\
\vspace{0mm}
\ey
$$
 } 
}
   i.e. the duplicator 
   copies its input to 
   two outputs, and 
\i the multiplier, which receives 
   on its input ports 
   those values that duplicator gives. 
\ei 

A process $Square$, 
corresponding to such a composition 
is defined as follows: 
$$\by Square\eam \\ \eam \b{
Dup[pass_1/Out_1, pass_2/Out_2]\;\mid\\\mid
Mul[pass_1/In_1, pass_2/In_2]}
\setminus\{pass_1,pass_2\}\ey$$

A flow graph of the process $Square$ 
has the form 
{\small
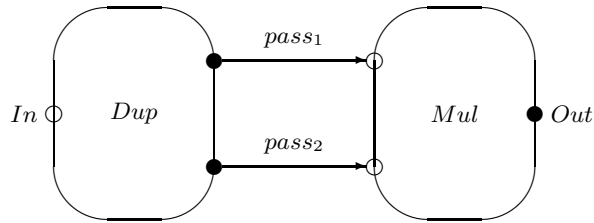
$$\by
\begin{picture}(0,100)

\put(60,50){\oval(60,80)}
\put(-60,50){\oval(60,80)}

\put(-90,50){\circle{6}}
\put(90,50){\circle*{6}}

\put(-30,70){\circle*{6}}
\put(-30,30){\circle*{6}}
\put(30,70){\circle{6}}
\put(30,30){\circle{6}}

\put(-27,70){\vector(1,0){54}}
\put(-27,30){\vector(1,0){54}}

\put(-60,50){\makebox(0,0)[c]{$Dup$}}
\put(60,50){\makebox(0,0)[c]{$Mul$}}
\put(-96,50){\makebox(0,0)[r]{$In$}}
\put(96,50){\makebox(0,0)[l]{$Out$}}

\put(0,75){\makebox(0,0)[b]{$pass_1$}}
\put(0,35){\makebox(0,0)[b]{$pass_2$}}

\end{picture}
\ey
$$ } 

However, the process $Square$ 
does not meet the specification 
$Square\_Spec$
(i.e. $Square$ and $Square\_Spec$ are not
observationally equivalent).
This fact is easy to detect
by a construction of a graph 
representation of $Square$, 
which, by definition 
of operations of parallel composition, 
restriction and renaming, 
is the following: 

{\def\arraystretch{1}
{\small
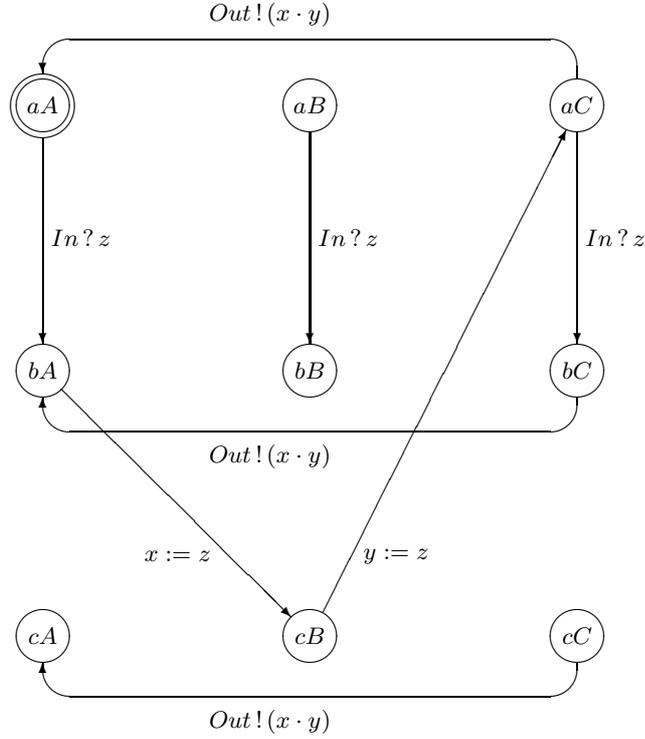
$$\by
\begin{picture}(210,240)

\put(0,200){\oval(20,20)}
\put(0,200){\oval(24,24)}
\put(0,200){\makebox(0,0)[c]{$aA$}}

\put(0,100){\oval(20,20)}
\put(0,100){\makebox(0,0)[c]{$bA$}}

\put(0,0){\oval(20,20)}
\put(0,0){\makebox(0,0)[c]{$cA$}}

\put(100,200){\oval(20,20)}
\put(100,200){\makebox(0,0)[c]{$aB$}}

\put(100,100){\oval(20,20)}
\put(100,100){\makebox(0,0)[c]{$bB$}}

\put(100,0){\oval(20,20)}
\put(100,0){\makebox(0,0)[c]{$cB$}}

\put(200,200){\oval(20,20)}
\put(200,200){\makebox(0,0)[c]{$aC$}}

\put(200,100){\oval(20,20)}
\put(200,100){\makebox(0,0)[c]{$bC$}}

\put(200,0){\oval(20,20)}
\put(200,0){\makebox(0,0)[c]{$cC$}}

\put(0,-13){\vector(0,1){3}}
\put(200,-13){\line(0,1){3}}
\put(10,-13){\oval(20,20)[bl]}
\put(190,-13){\oval(20,20)[br]}
\put(190,-23){\line(-1,0){180}}
\put(85,-28){\makebox(0,0)[t]{
$Out\,!\,(x\cdot y)$
}}

\put(0,87){\vector(0,1){3}}
\put(200,87){\line(0,1){3}}
\put(10,87){\oval(20,20)[bl]}
\put(190,87){\oval(20,20)[br]}
\put(190,77){\line(-1,0){180}}
\put(85,72){\makebox(0,0)[t]{
$Out\,!\,(x\cdot y)$
}}

\put(0,215){\vector(0,-1){3}}
\put(200,215){\line(0,-1){5}}
\put(10,215){\oval(20,20)[tl]}
\put(190,215){\oval(20,20)[tr]}
\put(190,225){\line(-1,0){180}}
\put(85,230){\makebox(0,0)[b]{
$Out\,!\,(x\cdot y)$
}}

\put(0,150){\makebox(0,0)[l]{
$In\,?\,z$
}}

\put(100,150){\makebox(0,0)[l]{
$In\,?\,z$
}}

\put(200,150){\makebox(0,0)[l]{
$In\,?\,z$
}}

\put(0,188){\vector(0,-1){78}}
\put(100,190){\vector(0,-1){80}}
\put(200,190){\vector(0,-1){80}}

\put(7,93){\vector(1,-1){86}}

\put(105,9){\vector(1,2){91}}

\put(66,30){\makebox(0,0)[r]{
$x:=z$
}}

\put(117,30){\makebox(0,0)[l]{
$y:=z$
}}

\end{picture}
\\
\vspace{15mm}
\ey
$$
 } 
}

After a simplification of this process we obtain 
the process

{\small
\be{sdfkjsadlfjasd1}\by
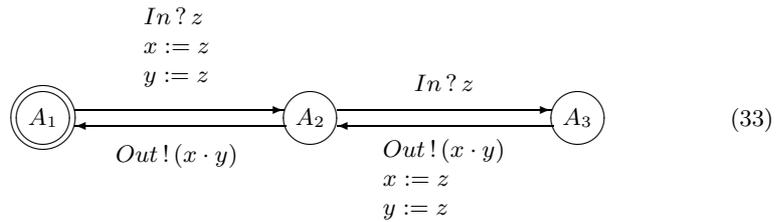
\begin{picture}(200,55)

\put(0,0){\oval(20,20)}
\put(0,0){\oval(24,24)}
\put(100,0){\oval(20,20)}
\put(200,0){\oval(20,20)}

\put(0,0){\makebox(0,0)[c]{$A_1$}}
\put(100,0){\makebox(0,0)[c]{$A_2$}}
\put(200,0){\makebox(0,0)[c]{$A_3$}}

\put(12,3){\vector(1,0){78}}
\put(91,-3){\vector(-1,0){79}}

\put(50,10){\makebox(0,0)[b]{$\by In\,?\,z\\ 
x:=z\\y:=z\ey$}}
\put(50,-10){\makebox(0,0)[t]{$Out\,!\,(x\cdot 
y)$}}

\put(110,3){\vector(1,0){80}}
\put(191,-3){\vector(-1,0){81}}

\put(150,10){\makebox(0,0)[b]{$In\,?\,z$}}
\put(150,-7){\makebox(0,0)[t]{$\by Out\,!\,
(x\cdot y)\\x:=z\\y:=z\ey$}}

\end{picture}
\\
\vspace{15mm}
\ey
\ee } 

which shows that 
\bi 
\i the process $Square$ can execute 
   two input actions together (i.e.
   without an execution of an output action
   between them), and\i the process 
   $Square\_Spec$ can not do that.
\ei

The process $Square$ meets another 
specification: 
$$Square\_Spec' \eam
\b{{\it Buf} [pass/Out]\;\mid\\\mid 
Square\_Spec[pass/In]}\setminus \{pass\}$$
where ${\it Buf}$ is a 
buffer which can store one message,
whose behavior is represented 
by the diagram 
{\small
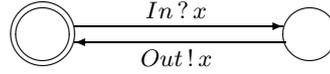
$$\by
\begin{picture}(100,30)

\put(0,0){\oval(20,20)}
\put(0,0){\oval(24,24)}
\put(100,0){\oval(20,20)}

\put(12,3){\vector(1,0){78}}
\put(91,-3){\vector(-1,0){79}}

\put(50,6){\makebox(0,0)[b]{$In\,?\,x$}}
\put(50,-6){\makebox(0,0)[t]{$Out\,!\,x$}}

\end{picture}
\\
\vspace{10mm}
\ey
$$ } 

A flow graph of $Square\_Spec'$ 
has the form
{\small
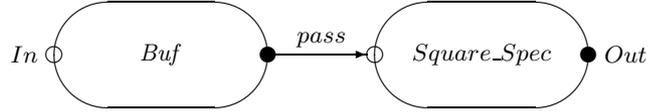
$$\by
\begin{picture}(0,50)

\put(60,20){\oval(80,40)}
\put(-60,20){\oval(80,40)}

\put(-100,20){\circle{6}}
\put(100,20){\circle*{6}}

\put(-20,20){\circle*{6}}
\put(20,20){\circle{6}}

\put(-17,20){\vector(1,0){34}}

\put(-60,20){\makebox(0,0)[c]{${\it Buf}$}}
\put(60,20){\makebox(0,0)[c]
{$Square\_Spec$}}
\put(-106,20){\makebox(0,0)[r]{$In$}}
\put(106,20){\makebox(0,0)[l]{$Out$}}

\put(0,23){\makebox(0,0)[b]{$pass$}}

\end{picture}
\ey
$$ } 

A simplified process $Square\_Spec'$
has the form

{\small
\be{sdfkjsadlfjasd2}\by
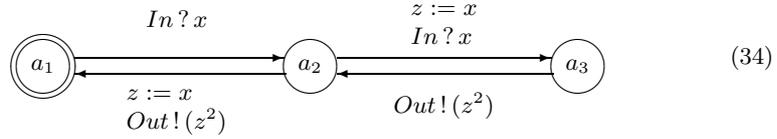
\begin{picture}(200,35)

\put(0,0){\oval(20,20)}
\put(0,0){\oval(24,24)}
\put(100,0){\oval(20,20)}
\put(200,0){\oval(20,20)}

\put(0,0){\makebox(0,0)[c]{$a_1$}}
\put(100,0){\makebox(0,0)[c]{$a_2$}}
\put(200,0){\makebox(0,0)[c]{$a_3$}}

\put(12,3){\vector(1,0){78}}
\put(91,-3){\vector(-1,0){79}}

\put(50,15){\makebox(0,0)[b]{$ In\,?\,x $}}
\put(50,-5){\makebox(0,0)[t]{$\by 
z:=x\\Out\,!\,(z^2)\ey$}}

\put(110,3){\vector(1,0){80}}
\put(191,-3){\vector(-1,0){81}}

\put(150,5){\makebox(0,0)[b]{$\by 
z:=x\\In\,?\,x\ey$}}
\put(150,-10){\makebox(0,0)[t]{$Out\,!\,(z^2)$}}

\end{picture}
\\
\vspace{5mm}
\ey
\ee } 

The statement that $Square$ meets the 
specification $Square\_Spec'$ 
can be formalized as 
\be{zfdgdsfgsdfg}
\re{sdfkjsadlfjasd1}\;\approx 
\re{sdfkjsadlfjasd2}\ee

We justify \re{zfdgdsfgsdfg} 
with use of theorem \ref{th1}. 
At first, we rename variables 
of the process \re{sdfkjsadlfjasd2}, 
i.e. instead of \re{sdfkjsadlfjasd2}
we shall consider the process
{\small
\be{sdfkjsadlfjasd3}\by
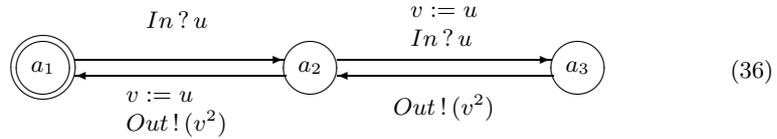
\begin{picture}(200,40)

\put(0,0){\oval(20,20)}
\put(0,0){\oval(24,24)}
\put(100,0){\oval(20,20)}
\put(200,0){\oval(20,20)}

\put(0,0){\makebox(0,0)[c]{$a_1$}}
\put(100,0){\makebox(0,0)[c]{$a_2$}}
\put(200,0){\makebox(0,0)[c]{$a_3$}}

\put(12,3){\vector(1,0){78}}
\put(91,-3){\vector(-1,0){79}}

\put(50,15){\makebox(0,0)[b]{$ In\,?\,u $}}
\put(50,-5){\makebox(0,0)[t]{$\by 
v:=u\\Out\,!\,(v^2)\ey$}}

\put(110,3){\vector(1,0){80}}
\put(191,-3){\vector(-1,0){81}}

\put(150,5){\makebox(0,0)[b]{$\by 
v:=u\\In\,?\,u\ey$}}
\put(150,-10){\makebox(0,0)[t]{$Out\,!\,
(v^2)$}}

\end{picture}
\\
\vspace{10mm}
\ey
\ee } 

To prove 
$\re{sdfkjsadlfjasd1}\;\approx \re{sdfkjsadlfjasd3}$
with use of  theorem \ref{th1}
we define the formulas 
$b_{A_i,a_j}$ (where $i,j=1,2,3$)
as 
follows: 
\bi
\i $b_{A_i,a_j}\eam \bot$, if $i\neq j$
\i $b_{A_1,a_1}\eam \top$
\i $b_{A_2,a_2}\eam (x=y=z=u)$
\i $b_{A_3,a_3}\eam (x=y=v)\wedge(z=u)$.
\ei


\end{document}